\documentclass[twocolumn]{emulateapj}

\begin{document}

\title{Observational and Dynamical Characterization of Main-Belt Comet P/2010 R2 (La Sagra)
        \footnote{
        Some data presented herein were obtained
        at the Isaac Newton Group of telescopes on La Palma as part of Program I/2010B/P14,
        at the Gemini Observatory under program GN-2011B-Q-17, with the Danish 1.54m telescope
        at the ESO La Silla Observatory, and at the W.\ M.
        Keck Observatory.  The Keck Observatory is operated as a scientific partnership
        among the California Institute of Technology, the University of
        California, and the National Aeronautics and Space Administration, and
        was made possible by the generous financial support
        of the W.\ M. Keck Foundation.  
        }
}

\author{
Henry H.\ Hsieh\altaffilmark{1,a},
Bin Yang\altaffilmark{1},~
Nader Haghighipour\altaffilmark{1},~
Bojan Novakovi\'c\altaffilmark{2},~
Robert Jedicke\altaffilmark{1},~
Richard J.\ Wainscoat\altaffilmark{1},~
Larry Denneau\altaffilmark{1},~
Shinsuke Abe\altaffilmark{3},~
Wen-Ping Chen\altaffilmark{3},~
Alan Fitzsimmons\altaffilmark{4},~
Mikael Granvik\altaffilmark{5},~
Tommy Grav\altaffilmark{6},~
Wing Ip\altaffilmark{3},~
Heather M.\ Kaluna\altaffilmark{1},~
Daisuke Kinoshita\altaffilmark{3},~
Jan Kleyna\altaffilmark{1},~
Matthew M.\ Knight\altaffilmark{7},~
Pedro Lacerda\altaffilmark{4,b},~
Carey M.\ Lisse\altaffilmark{8},~
Eric Maclennan\altaffilmark{9},~
Karen J.\ Meech\altaffilmark{1},~
Marco Micheli\altaffilmark{1},~
Andrea Milani\altaffilmark{10},~
Jana Pittichov\'a\altaffilmark{1},~
Eva Schunova\altaffilmark{1,11},~
David J.\ Tholen\altaffilmark{1},~
Lawrence H.\ Wasserman\altaffilmark{7},~
William S.\ Burgett\altaffilmark{1},~
K.~C.\ Chambers\altaffilmark{1},~
Jim N.\ Heasley\altaffilmark{1},~
Nick Kaiser\altaffilmark{1},~
Eugene A.\ Magnier\altaffilmark{1},~
Jeffrey S.\ Morgan\altaffilmark{1},~
Paul A.\ Price\altaffilmark{12},~
Uffe G.\ J{\o}rgensen\altaffilmark{13,14},~
Martin Dominik\altaffilmark{15,c},~
Tobias Hinse\altaffilmark{16},~
Kailash Sahu\altaffilmark{17},~
and
Colin Snodgrass\altaffilmark{18},~
}


\altaffiltext{1}{Institute for Astronomy, University of Hawaii, 2680 Woodlawn Drive, Honolulu HI 96822, USA}
\altaffiltext{2}{Department of Astronomy, Faculty of Mathematics, University of Belgrade, Studentski trg 16, 11000 Belgrade, Serbia}
\altaffiltext{3}{National Central University, 300 Jhongda Rd, Jhongli City, Taoyuan 320, Taiwan}
\altaffiltext{4}{Astronomy Research Centre, Queens University Belfast, Belfast BT7 1NN, United Kingdom}
\altaffiltext{5}{Department of Physics, P.O. Box 64, 00014 University of Helsinki, Finland}
\altaffiltext{6}{Department of Physics and Astronomy, Johns Hopkins University, 3400 North Charles Street, Baltimore, MD 21218, USA}
\altaffiltext{7}{Lowell Observatory, 1400 West Mars Hill Road, Flagstaff, AZ 86001, USA}
\altaffiltext{8}{Planetary Exploration Group, Space Department, Johns Hopkins University Applied Physics Laboratory, Laurel, MD 20723, USA}
\altaffiltext{9}{Northern Arizona University}
\altaffiltext{10}{Dipartimento di Matematica, Universit{\`a} di Pisa, Largo Pontecorvo 5, 56127 Pisa, Italy}
\altaffiltext{11}{Department of Astronomy, Physics of the Earth and Meteorology, Comenius University, Mlynska dolina, 842 48 Bratislava, Slovakia}
\altaffiltext{12}{Department of Astrophysical Sciences, Princeton University, Princeton, NJ 08544, USA}
\altaffiltext{13}{Niels Bohr Institute, University of Copenhagen, Juliane Maries Vej 30, 2100 Copenhagen, Denmark}
\altaffiltext{14}{Centre for Star and Planet Formation, Geological Museum, {\O}ster Voldgade 5, 1350 Copenhagen, Denmark}
\altaffiltext{15}{SUPA, University of St Andrews, School of Physics \& Astronomy, North Haugh, St Andrews, KY16 9SS, United Kingdom}
\altaffiltext{16}{Korea Astronomy \& Space Science Institute (KASI), 776 Daedukdae-ro, Yuseong-gu, Daejeon, Republic of Korea}
\altaffiltext{17}{Space Telescope Science Institute, 3700 San Martin Drive, Baltimore, MD 21218, USA}
\altaffiltext{18}{Max-Planck-Institut f\"ur Sonnensystemforschung, 37191 Katlenburg-Lindau, Germany}
\altaffiltext{a}{Hubble Fellow}
\altaffiltext{b}{Michael West Fellow}
\altaffiltext{c}{Royal Society University Research Fellow}

\email{hsieh@ifa.hawaii.edu}

\slugcomment{Submitted, 2011-09-27; Accepted, 2012-02-14}

\begin{abstract}
We present observations of the recently discovered comet-like main-belt object P/2010 R2 (La Sagra) obtained by Pan-STARRS 1 and the Faulkes Telescope-North on Haleakala in Hawaii, the University of Hawaii 2.2~m, Gemini-North, and Keck I telescopes on Mauna Kea, the Danish 1.54~m telescope (operated by the MINDSTEp consortium) at La Silla, and the Isaac Newton Telescope on La Palma.  An antisolar dust tail is observed to be present from August 2010 through February 2011, while a dust trail aligned with the object's orbit plane is also observed from December 2010 through August 2011.  Assuming typical phase darkening behavior, P/La Sagra is seen to increase in brightness by $>1$~mag between August 2010 and December 2010, suggesting that dust production is ongoing over this period.  These results strongly suggest that the observed activity is cometary in nature ({\it i.e.}, driven by the sublimation of volatile material), and that P/La Sagra is therefore the most recent main-belt comet to be discovered.  We find an approximate absolute magnitude for the nucleus of $H_R=17.9\pm0.2$~mag, corresponding to a nucleus radius of $\sim0.7$~km, assuming an albedo of $p=0.05$.  Comparing the observed scattering surface areas of the dust coma to that of the nucleus when P/La Sagra was active, we find dust-to-nucleus area ratios of $A_d/A_N=30-60$, comparable to those computed for fellow main-belt comets 238P/Read and P/2008 R1 (Garradd), and one to two orders of magnitude larger than for two other main-belt comets (133P/Elst-Pizarro and 176P/LINEAR).  Using optical spectroscopy to search for CN emission, we do not detect any conclusive evidence of sublimation products (i.e., gas emission), finding an upper limit CN production rate of $Q_{\rm CN}<6\times10^{23}$~mol~s$^{-1}$, from which we infer an H$_2$O production rate of $Q_{\rm H_2O}<10^{26}$~mol~s$^{-1}$. Numerical simulations indicate that P/La Sagra is dynamically stable for $>100$~Myr, suggesting that it is likely native to its current location and that its composition is likely representative of other objects in the same region of the main belt, though the relatively close proximity of the 13:6 mean-motion resonance with Jupiter and the (3,-2,-1) three-body mean-motion resonance with Jupiter and Saturn mean that dynamical instability on larger timescales cannot be ruled out.
\end{abstract}


\newpage

\section{INTRODUCTION}

Comet P/2010 R2 (La Sagra), hereafter P/La Sagra, was discovered by J.\ Nomen in images obtained on 2010 September 14.9 (UT) using the 0.45~m La Sagra Observatory in southern Spain \citep{nom10}.  With a semimajor axis of $a=3.099$~AU, eccentricity of $e=0.154$, inclination of $i=21.39\degr$, and a Tisserand parameter (with respect to Jupiter) of $T_J=3.099$, it was immediately suspected to be a member of the recently identified class of main-belt comets \citep[MBCs;][]{hsi06}, which are objects that exhibit cometary activity likely due to the sublimation of volatile ice, yet are dynamically indistinguishable from main-belt asteroids.  Prior to the discovery of P/La Sagra, four MBCs --- 133P/Elst-Pizarro, 176P/LINEAR, 238P/Read, and P/2008 R1 (Garradd) --- were recognized, while a sixth possible MBC (P/2006~VW$_{139}$) has just recently been identified \citep{hsi11c}.  Two other comet-like main-belt objects --- P/2010 A2 (LINEAR) and (596) Scheila --- have also been observed, but as their comet-like morphologies are believed to be due to impact-generated ejecta clouds \citep{jew10,jew11,sno10,bod11,yan11,hai12}, and not cometary (i.e., sublimation-driven) dust emission, they are better characterized as disrupted asteroids.

Observationally distinguishing MBCs and disrupted asteroids is not straightforward as both types of objects exhibit visible dust emission in the form of comet-like features like comae and tails.  Distinctive morphological features helped betray the true natures of P/2010~A2 and Scheila, however \citep[][and references within]{hsi12a}.  In the case of P/2010 A2, a gap between what appeared to be the nucleus of the ``comet'' and the dust tail indicated that the observed dust emission was possibly an impulsively-generated ejecta cloud that produced by an impact onto P/2010 A2's surface and was then drifting away.  This hypothesis was later corroborated by numerical dust modeling and other physical arguments \citep{jew10,sno10}.  In the case of Scheila, no gap was seen between the nucleus and its emitted dust, but instead of forming a single tail, the dust emitted from Scheila appeared to form multiple plumes.  \citet{hsi12a} considered various scenarios that could produce multiple dust plumes, finding scenarios for sublimation-driven dust emission to be physically implausible as they required multiple active sites, which were argued to be unlikely on Scheila.  Instead, a simpler scenario in which a single oblique impact caused dust emission in the form of a hollow cone which was then pushed back in the antisolar direction by radiation pressure was favored \citep{ish11a,ish11b}.  In this scenario, what appeared to be multiple plumes of dust were proposed to simply be optical projection effects (i.e., limb brightening of the hollow dust structure).

Early observation reports for P/La Sagra indicated that it exhibited a largely classical cometary morphology, with a visible coma and tail \citep{nom10}, suggesting that it could be a true comet \citep{hsi12a}, and was therefore likely to be just the fifth MBC to be discovered at that time.  Numerical modeling by \citet{mor11} furthermore showed that the evolution of P/La Sagra's dust tail was consistent with sublimation-driven dust emission, though they did not explicitly test impulsive impact-driven emission scenarios.  In this paper, we seek to further assess whether P/La Sagra's dust emission is likely to be cometary (sublimation-driven) in nature, or if the object could instead be a disrupted asteroid.  We also conduct observational and dynamical characterization analyses of the object to place it in the context of the small but growing known population of comet-like objects in the main asteroid belt.

\section{OBSERVATIONS}
\label{obs}


Imaging observations of P/La Sagra were made in photometric conditions on multiple occasions in 2010 and 2011 (Table~\ref{obslog}) by the 2.0~m Faulkes Telescope North (FTN) on Haleakala in Hawaii, the University of Hawaii (UH) 2.2~m telescope, the 8~m Gemini-North Observatory, and the 10~m Keck I telescope, all on Mauna Kea in Hawaii, the 1.54~m Danish telescope at La Silla Observatory in Chile, and the 2.5~m Isaac Newton Telescope (INT) at the Roque de los Muchachos Observatory on La Palma in the Canary Islands (Spain).  In addition, precovery observations of P/La Sagra were also obtained with the 1.8~m Pan-STARRS 1 (PS1) survey telescope on Haleakala.  To ensure proper calibration of PS1 data, we also obtained calibration observations using the Lowell Observatory 31$''$ telescope in Flagstaff, Arizona.

PS1 is a wide-field synoptic survey telescope that employs a 1.4 gigapixel camera consisting of a mosaic of 60 orthogonal transfer arrays (OTAs), each consisting of 64 $590\times598$~pixel CCDs, giving a total field of view 3.2 degrees in diameter (image scale of $0\farcs26$~pixel$^{-1}$).  PS1 observations were made using a Sloan Digital Sky Survey (SDSS) $r'$-band-like filter designated $r_{\rm P1}$ \citep{stu10}.
Gemini observations (obtained as part of program GN-2011B-Q-17) were made using the Gemini Multi-Object Spectrograph \citep[GMOS;][]{hoo04} in imaging mode.  GMOS employs three $2048\times2048$ EEV CCDs behind a SDSS-like $r'$-band filter, and has an image scale of $0\farcs1456$~pixel$^{-1}$ when using $2\times2$ binning as we did for our observations of P/La Sagra.
Keck observations were made using the Low Resolution Imaging Spectrometer \citep[LRIS;][]{oke95} in both imaging and spectroscopic mode.  LRIS employs a Tektronix $2048 \times 2048$ CCD with an image scale of $0\farcs135$~pixel$^{-1}$ and standard Kron-Cousins $B$-band and $R$-band filters.
The Danish 1.54~m telescope is currently exclusively used by the Microlensing Network for the Detection of Small Terrestrial Exoplanets (MiNDSTEp) consortium.
Observations on the Danish 1.54~m were made using the Danish Faint Object Spectrograph and Camera (DFOSC) focal-reducing imager behind a standard Kron-Cousins $R$-band filter.  The DFOSC imager has pixel dimensions of $2102\times2102$ and a pixel scale of $0\farcs39$~pixel$^{-1}$.
FTN observations were made using a $1024\times1024$ e2v CCD with an image scale of $0\farcs2785$~pixel$^{-1}$ behind a Bessell $R$-band filter.

Except for observations with PS1, FTN, and the INT, all other observations were conducted using non-sidereal tracking at the apparent rate and direction of motion of P/La Sagra on the sky.  Sidereal tracking was used by PS1 because this was the standard mode of operation for survey operations.  FTN observations were obtained using its web-based remote control interface, through which non-sidereal tracking is not possible.  The INT is normally capable of tracking solar system objects non-sidereally, but was unable to do so during our observing run due to mechanical problems.

We performed standard image preparation (bias subtraction and flat-field reduction) for UH 2.2~m, Gemini, Keck, Danish 1.54~m, and INT data using Image Reduction and Analysis Facility (IRAF) software.  For Keck data, flat fields were constructed from images of the illuminated interior of the Keck I dome, while for the other telescopes, dithered images of the twilight sky were used to construct flat fields.  PS1 data were reduced using the system's Image Processing Pipeline \citep[IPP;][]{mag06} and then calibrated using field stars imaged later by the Lowell 31$''$ telescope.  Calibration of Gemini data was performed using field star magnitudes provided by the Sloan Digital Sky Survey \citep[SDSS;][]{yor00}.  Photometry of \citet{lan92} standard stars and field stars was performed for all data using IRAF and obtained by measuring net fluxes within circular apertures, with background sampled from surrounding circular annuli.  Conversion of magnitudes measured from PS1 and Gemini data obtained using SDSS-like $r'$-band filters to their Kron-Cousins $R$-band equivalents was accomplished using the transformation equations derived by R.\ Lupton and made available on the SDSS website ({\tt http://www.sdss.org/}).  Comet photometry was performed using both rectangular and circular apertures as described below, where to avoid dust contamination from the comet itself, background sky statistics are measured manually in regions of blank sky near, but not adjacent, to the object.  Several (5 to 10) field stars in the comet images were also measured to correct for any extinction variation (typically negligible) during each night.

In addition to imaging observations, we also secured optical spectra of P/La Sagra on 2010 October 05 with LRIS on Keck. We adopted LRIS's $1\farcs0$-wide long-slit mask, the 400/3400 grism, and the 460 dichroic on the blue side, giving a dispersion of 1.08 \AA~pixel$^{-1}$ and a spectral resolution of approximately 7 \AA. A total of 3600~s of data were obtained from four integrations on the comet while it was at an airmass of $\sim$1.1. Unfortunately, when taking calibration images, our observations were interrupted due to fog and we were forced to close the dome for over three hours. As such, we were only able to record a spectrum of a flux standard star and failed to obtain any spectra of nearby G-type standard stars, complicating our reduction and analysis (Section~\ref{specreduction}). Data reduction was accomplished using IRAF.

\section{RESULTS \& ANALYSIS\label{results}}

\subsection{Morphological Analysis\label{imageanalysis}}

\begin{figure}
\plotone{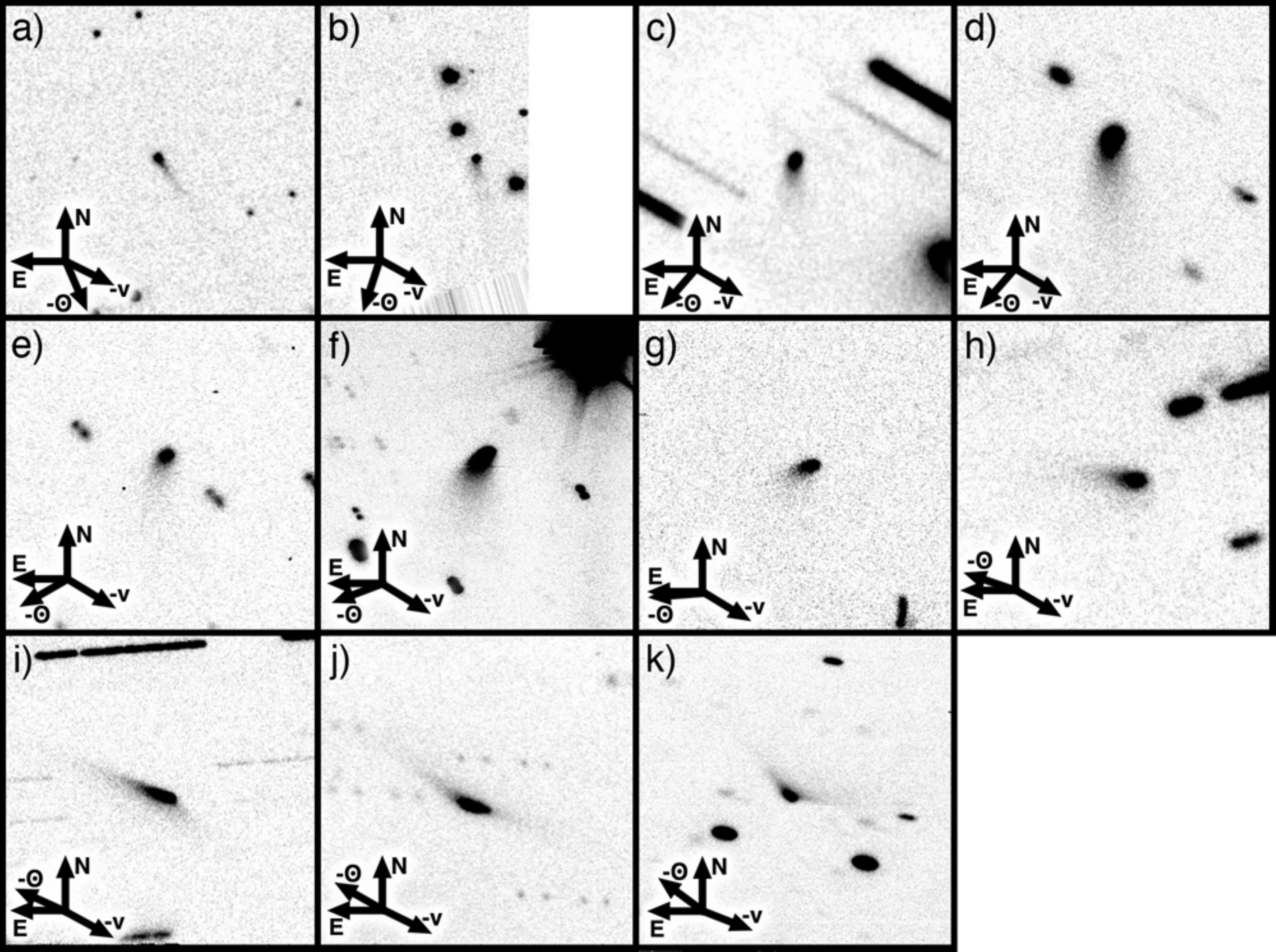}
\caption{\small Composite images of P/La Sagra (at the center of each panel) constructed
from data obtained on
(a) 2010 August 16 (80~s of effective exposure time on PS1 in $r'$-band),
(b) 2010 September 08 (40~s on PS1 in $r'$-band),
(c) 2010 September 19 (2400~s on the Danish 1.54~m in $R$-band),
(d) 2010 September 19 (1500~s on the UH 2.2~m in $R$-band),
(e) 2010 September 29 (450~s on FTN in $R$-band),
(f) 2010 October 05 (240~s on Keck I in $R$-band),
(g) 2010 October 19 (900~s on the UH 2.2~m in $R$-band),
(h) 2010 November 26 (1800~s on the UH 2.2~m in $R$-band),
(i) 2010 December 12 (5400~s on the UH 2.2~m in $R$-band),
(j) 2010 December 31 (1200~s on the INT in $R$-band), and
(k) 2011 February 03 (360~s on Keck I in $R$-band).
All panels are $60''\times60''$ in size with north (N), east (E), the antisolar direction ($-\odot$),
and the negative heliocentric velocity vector ($-v$), as projected on the sky, marked.
}
\label{images1}
\end{figure}

\begin{figure}
\plotone{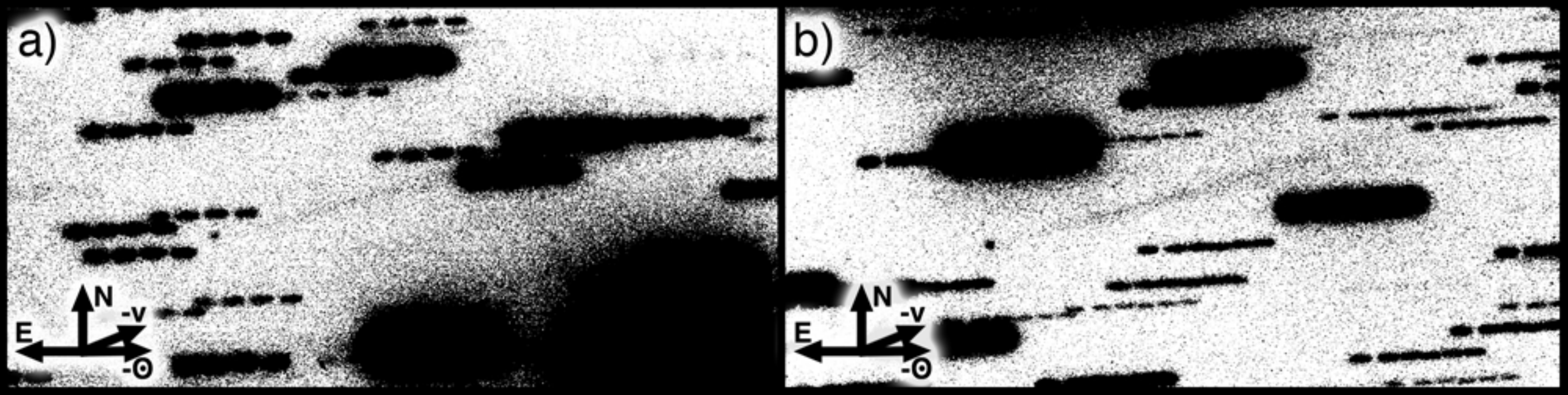}
\caption{\small Composite images of P/La Sagra (point-like object in the lower left of each panel) constructed
from data obtained on
(a) 2011 August 26 (1200~s on Keck I in $R$-band) and
(b) 2011 August 31 (1800~s on Gemini in $r'$-band).
Panels are $120''\times60''$, with north (N), east (E), the antisolar direction ($-\odot$),
and the negative heliocentric velocity vector ($-v$), as projected on the sky, marked.
}
\label{images2}
\end{figure}

For each night during which observations were obtained, we construct composite images by aligning each night's set of images on the object's photocenter and adding them together. We see that P/La Sagra is clearly morphologically cometary between August 2010 and February 2011 (Figure~\ref{images1}), with a dust tail extending up to $\sim30''$ ($\sim4\times10^4$~km at the geocentric distance of the object at the time) from the nucleus's photocenter.  The projected orientation of this dust tail on the plane of the sky is seen to rotate counterclockwise throughout this period, lagging slightly behind the similarly rotating antisolar vector (as projected on the sky).  From August 2010 through November 2010, we observe a single dust tail aligned with the projection of the antisolar vector in the sky, while between December 2010 and February 2011, an apparent dust trail aligned with the projection of the heliocentric velocity vector (i.e., the plane of the object's orbit) appears in addition to the antisolar dust tail.  By August 2011 (Figure~\ref{images2}), active dust emission appears to have ceased, though a residual dust trail (presumably consisting of large dust particles ejected during earlier activity) aligned with the projection of the object's heliocentric velocity vector in the sky is observed extending at least $\sim$1~arcmin ($\sim1.5\times10^5$~km at the geocentric distance of the object at the time) from the nucleus.  This dust trail peaks in brightness well away ($\sim20''$) from the nucleus.

In the formulation of dust dynamics outlined by \citet{fin68}, the trajectory of an emitted dust particle depends on the orbit of the emitting comet, the particle's ejection time, its ejection velocity, and the ratio, $\beta$, of solar radiation pressure, $F_{\rm rad}$, to solar gravity, $F_{\rm grav}$, on the particle.  In cases where $\beta$ is small (i.e., for massive particles), radiation pressure is negligible, causing particles follow orbits very close that of the comet.  Over time, these large grains will spread out along the comet's orbit to form a potentially observable trail.  In cases where $\beta$ is large (i.e., for small particles), radiation pressure is dominant over solar gravity, giving particles nearly directly antisolar motion in the rest frame of the comet.  In the case of P/La Sagra, we observe both limiting cases:  a tail consistently oriented in the antisolar direction (likely composed of small, high-$\beta$ particles) from August 2010 through February 2011 (Figure~\ref{images1}), and a trail confined to the orbit plane (likely composed of large, low-$\beta$ particles) from December 2010 through August 2011 (Figures~\ref{images1} and \ref{images2}).  This assessment is in agreement with the modeling results obtained by \citet{mor11} who found dust grain sizes ranging from $\sim$5~$\mu$m to $\sim$1~cm.  Similar dust structures comprising widely-varying particle sizes have also been observed for comets 2P/Encke \citep{rea00,lis04} and C/Austin 1990 V \citep{lis98}.

Due to the ease by which they are accelerated by radiation pressure, the small particles comprising an antisolar tail are expected to dissipate quickly \citep[on timescales of several weeks at the distance of the main belt; cf.][]{hsi04} unless they are continually replenished.  No such separation of the antisolar tail from the nucleus is observed at any time from August 2010 through February 2011, a span of 5.5 months, strongly suggesting that continuous dust production was ongoing over this time period, consistent with cometary activity.  We note that the persistence of the orbit-aligned dust trail until August 2011 does not mean that dust production continued up until this time as well.  In this latter case, the longevity of the dust trail is instead likely due to the large sizes and therefore slow dissipation rates of the particles in the trail \citep[cf.\ P/2010 A2;][]{jew10,sno10}, not continuous replenishment.

Qualitatively, we find that P/La Sagra's morphological appearance and evolution strongly suggest that its dust emission is cometary in nature.  We caution however that more detailed dust modeling \citep[similar that performed by][but also considering impulsive emission scenarios]{mor11} will be required for quantitative confirmation of the conclusions reached here.

\subsection{Photometric Analysis\label{photometry}}

\subsubsection{Nucleus Size, Shape, and Color\label{nucleussize}}

Vigorous activity, such as that exhibited by P/La Sagra during the majority of our 2010 and 2011 observations, normally interferes with the measurement of physical properties for cometary nuclei (e.g., size, rotation period, and phase darkening behavior) due to obscuration caused by coma when such activity is present.  As such, in order to accurately measure the physical properties of comet nuclei, it is typically necessary to observe them once activity has mostly or entirely ceased, as we did with P/La Sagra in August 2011 using the Keck I and Gemini telescopes (Table~\ref{obslog}). We measure a mean apparent $R$-band magnitude (averaged in flux space and weighted by nominal uncertainties) of $m=23.9\pm0.1$~mag, with a magnitude variation between the two nights of $\Delta m=0.7\pm0.1$~mag ($m_{\rm avg}=24.1\pm0.1$~mag on 2011 August 26, and $m_{\rm avg}=23.4\pm0.1$~mag on 2011 Aug 31).  Provided that the photometric variation between these two nights is real (intra-night variations are on the order of $\sim$0.2~mag, with observations on each night only spanning 30~min or less) and due to rotation, the minimum ratio of the axes of the body as projected in the plane of the sky is then given by
\begin{equation}
a_N/b_N = 10^{0.4\Delta m}
\end{equation}
giving us a lower limit (due to unknown projection effects) to the axis ratio for P/La Sagra of $a_N/b_N=1.9\pm0.3$.  Given the extremely limited data set on which this calculation is based, this axis ratio should of course be considered as a preliminary result only.  Better constraints on the shape and structure of the nucleus of P/La Sagra will require additional observational sampling of the object's lightcurve (likely requiring large telescopes, given the small size of the object), ideally at multiple observing geometries.

Using the range of slope parameter ($G$) values in the IAU $H,G$ photometric system computed for fellow MBCs 133P, 176P, 238P, and P/Garradd \citep[$-0.03<G_R<0.26$;][]{hsi09a,hsi11b,mac12}, we find an estimated absolute magnitude of $H_R=17.9\pm0.2$~mag.  Adopting a $R$-band albedo of $p_R=0.05$ \citep[as measured for 133P and 176P;][]{hsi09a}, we then find an approximate effective nucleus radius of $r_N=0.7\pm0.1$~km, making it comparable in size to 238P, about three times larger than P/Garradd, and about three times smaller than 133P and 176P (Table~\ref{orbelems}).  Using the approximate axis ratio derived above and assuming a triaxial ellipsoidal shape, we find semimajor and semiminor axes for the nucleus of $a_N=1.0$~km and $b_N=0.5$~km, respectively.

We note that photometric variations on the scale of $\Delta m=0.7$~mag were not seen at any time while P/La Sagra was active (August 2010 through February 2011) for any observations that were comparably closely spaced in time as the August 2011 observations analyzed above.  For example, observations on four separate occasions between 2010 Sep 08 and Sep 29 exhibited a maximum variation of just $\sim0.1$~mag (Table~\ref{photresults}).  This lack of corroborating detections of significant photometric variation during that period does not invalidate our detection of a $\Delta m=0.7$~mag variation in August 2011, however, since rotational variations for a nucleus obscured by a steady-state coma (assumed to be optically thin) will be effectively damped by that coma.  The extent of this damping can be determined using
\begin{equation}
\Delta m_{\rm obs}  = 2.5\log{F_{\rm obs,max}\over F_{\rm obs,min}} = 2.5\log\left({F_{N,{\rm max}}+F_d\over F_{N,{\rm min}}+F_d}\right)
\label{damping}
\end{equation}
where $\Delta m_{\rm obs}$ is the observed photometric range, $F_{\rm obs,max}$ and $F_{\rm obs,min}$ are the maximum and minimum fluxes observed for the active comet, $F_{N,{\rm max}}$ and $F_{N,{\rm min}}$ are the maximum and minimum fluxes for which the nucleus is responsible, and the flux due to dust, $F_d$, is assumed to be constant \citep{hsi11a}.

Assuming that the nucleus and dust grains have similar albedos, flux terms in Equation~\ref{damping} can be considered equivalent to the corresponding scattering surface areas of the comet components being considered.  We find in the following section (Section~\ref{activityevolution}) that the ratio of scattering surface areas of dust and the nucleus during the four September observations discussed above was $A_d/A_N\sim30$ (Table~\ref{photresults}).  Thus, using $A_{N,{\rm max}}=\pi a_N r_N$, $A_{N,{\rm min}}=\pi b_N r_N$, and $A_d=\pi r_N^2$, we find $\Delta m_{\rm obs} = 0.025$~mag.  In other words, during this period, even the minor $\sim0.1$~mag photometric variation we observed for the object is unlikely to be due to the rotation of the nucleus.  Instead, we surmise that the observed variation is more likely due to fluctuations in the seeing conditions between the different nights, causing fluctuations in the amount of coma contained within the fixed $5\farcs0$ photometry apertures we used to measure the near-nucleus flux of the comet in all of our images.  We also conclude that our non-detection of significant photometric variations in data obtained when P/La Sagra was active is not inconsistent with our apparent detection of a significant photometric variation while the object was inactive.

Multi-filter observations using LRIS on Keck I also permitted us to measure a $B-R=0.9\pm0.1$~mag color for P/La Sagra's inactive nucleus, which within our uncertainties, is consistent with the color of the Sun \citep[$B-R=1.03$~mag;][]{har82,har90}.  The colors we measure for P/La Sagra's dust coma while it was active are also approximately solar ($B-R=1.02\pm0.05$~mag on 2010 October 5 and $B-R=0.94\pm0.05$~mag on 2011 Feb 3), in good agreement with colors measured for the other MBCs \citep{jew09,hsi09b,hsi10b,hsi11a}.

\subsubsection{Activity Strength\label{activityevolution}}

Using the nucleus size computed above in Section~\ref{nucleussize}, the expected apparent nucleus magnitude for P/La Sagra at the time of our first observations in August 2010 is $m_R=22.2$~mag, not $m_R=18.49$~mag as measured (Table~\ref{photresults}). This discrepancy indicates that there was significant near-nucleus dust contamination even during our earliest observations of P/La Sagra.  Assuming that the phase behavior of P/La Sagra's dust coma is similar to that of the nuclei of other MBCs whose phase functions have been measured (cf.\ Section~\ref{nucleussize}; since we lack other meaningful constraints), we calculate the comet's corresponding absolute magnitudes, $m_{\rm tot}(1,1,0)$, at heliocentric and geocentric distances of $R=\Delta=1$~AU and phase angles of $\alpha=0\degr$ for each set of photometric measurements (using photometry apertures with radii of $5\farcs0$) of the nucleus (Table~\ref{photresults}).  By comparing these absolute magnitudes to our estimated absolute magnitude for the nucleus, we then compute the ratio of the scattering surface area of dust in the coma, $A_d$, to that of the nucleus, $A_N$, using
\begin{equation}
{A_d\over A_N} = {1-10^{0.4(m_{\rm tot}(1,1,0)-H_R)}\over 10^{0.4(m_{\rm tot}(1,1,0)-H_R)}}
\end{equation}
Then, assuming $A_N\sim\pi r_N^2$ and $a=10$~$\mu$m-radius grains with bulk densities of $\rho=1300$~kg~m$^{-3}$ \citep[cf.][]{hsi04,jut08}, we estimate the total dust mass, $M_d$, in the coma using
\begin{equation}
M_d = {4\over3}\rho a \cdot A_N\left({A_d\over A_N}\right)
\end{equation}
We list the values we compute for $A_d/A_N$ and $M_d$ using the above procedures in Table~\ref{photresults}.  Comparing the minimum and maximum values of $M_d$ during our observations, we find that P/La Sagra's coma increases in mass at an average net rate of $\dot M_d\sim0.1$~kg~s$^{-1}$ (likely precise to an order of magnitude, at best) between August and December 2010.  This net increase in total dust mass represents the sum of the positive contribution of new dust production from the nucleus and the negative contribution of dust dissipation as individual grains disperse beyond the photometry aperture.  As such, it represents a lower limit to the true dust production rate of the nucleus.

In addition to performing nucleus photometry, we also measure the total flux from the comet in our composite images from each night.  We do so by using rectangular photometry apertures enclosing the entire visible dust cloud and oriented to avoid field star contamination. Background sky levels were then measured from nearby areas of blank sky and subtracted to obtain net fluxes, which were then calibrated using standard stars to obtain absolute photometry ($m_{\rm tot}$).  Unfortunately, the faintness of the residual dust trail observed on 2011 August 26 and 31, and extensive contamination by nearby bright field stars means that we are unable to perform these measurements for data obtained on those dates.  Repeating the mass production rate analysis described above, we find that P/La Sagra's coma and tail increase in mass at about twice the rate measured for the near-nucleus coma, or an average net rate of $\dot M_d\sim0.2$~kg~s$^{-1}$.

Since this analysis does not rely on a fixed aperture size, this net mass loss rate is closer to the true dust production rate of the nucleus than that computed above, though still represents a lower limit due to the unknown dissipation rate of the dust as individual grains disperse from the coma and tail and become undetectable above the sky background.  Comparing with other MBCs, we nonetheless find that P/La Sagra has a dust production rate comparable to or larger than that of 238P \citep{hsi09b}, and at least an order of magnitude larger than those of 133P and 176P \citep{hsi04,hsi11a}.

For reference, we also use our measurements of the near-nucleus flux of the comet (using $5\farcs0$ photometry apertures) in each set of observations to estimate the dust contribution as parameterized by $Af\rho$ \citep{ahe84}.  These results are also shown in Table~\ref{photresults}, and again indicate significantly stronger activity for P/La Sagra as compared to 133P, for which $Af\rho$ was also systematically tabulated \citep{hsi10b}.

We also attempt to reproduce our observed photometry using a sublimation model similar to that used by \citep{hsi11b} to analyze the activity of 238P.  This effort was hampered, however, by our lack of observational constraints on the start time of activity and the rate of its eventual decline.  We do find that steadily increasing activity during the post-perihelion portion of P/La Sagra's orbit cannot be modeled by surface ice sublimation, indicating that the ice must instead reside at some depth below the surface of the nucleus, similar to what was found for 238P \citep{hsi11b}.  Unfortunately, the numerous free parameters that can be modified in our model mean that this finding is the only meaningful result that can be derived at this time.  Better sampling of the activity over a longer time baseline will be required to gain additional insight into the nature of P/La Sagra's activity via this method, and as such, should be considered an observational priority during its next expected period of activity (likely beginning in mid-2015).

We note that if P/La Sagra exhibits activity over a period of $\sim1$~year during each of its perihelion passages \citep[similar to 133P;][]{hsi04}, assuming an average mass loss rate of $\dot M_d=0.2$~kg~s$^{-1}$, a total mass loss of $\sim10^7$~kg per orbit would be expected.  Assuming a nucleus size of $r_N=0.7$~km (Section~\ref{nucleussize}) and $\rho=1300$~kg~m$^{-3}$, such a mass loss rate should consume the entire nucleus over the next $\sim10^5$~orbits, or $<10^6$~yr.  However, activity in MBCs (even over just a portion of their orbits) is expected to be transient \citep[cf.][]{hsi04}, perhaps amounting to just 1\% of an object's total lifetime \citep{hsi09}.  As such, we expect a timescale for nucleus disintegration due to cometary activity closer to $\sim10^8$~yr \citep{hsi09}, or approximately comparable to the $\sim10^8$-$10^9$~yr collisional lifetime expected for a main-belt asteroid the size of P/La Sagra's nucleus \citep{bot05}.

\subsubsection{Comparison to Other MBCs\label{actcomparison}}

\begin{figure}
\plotone{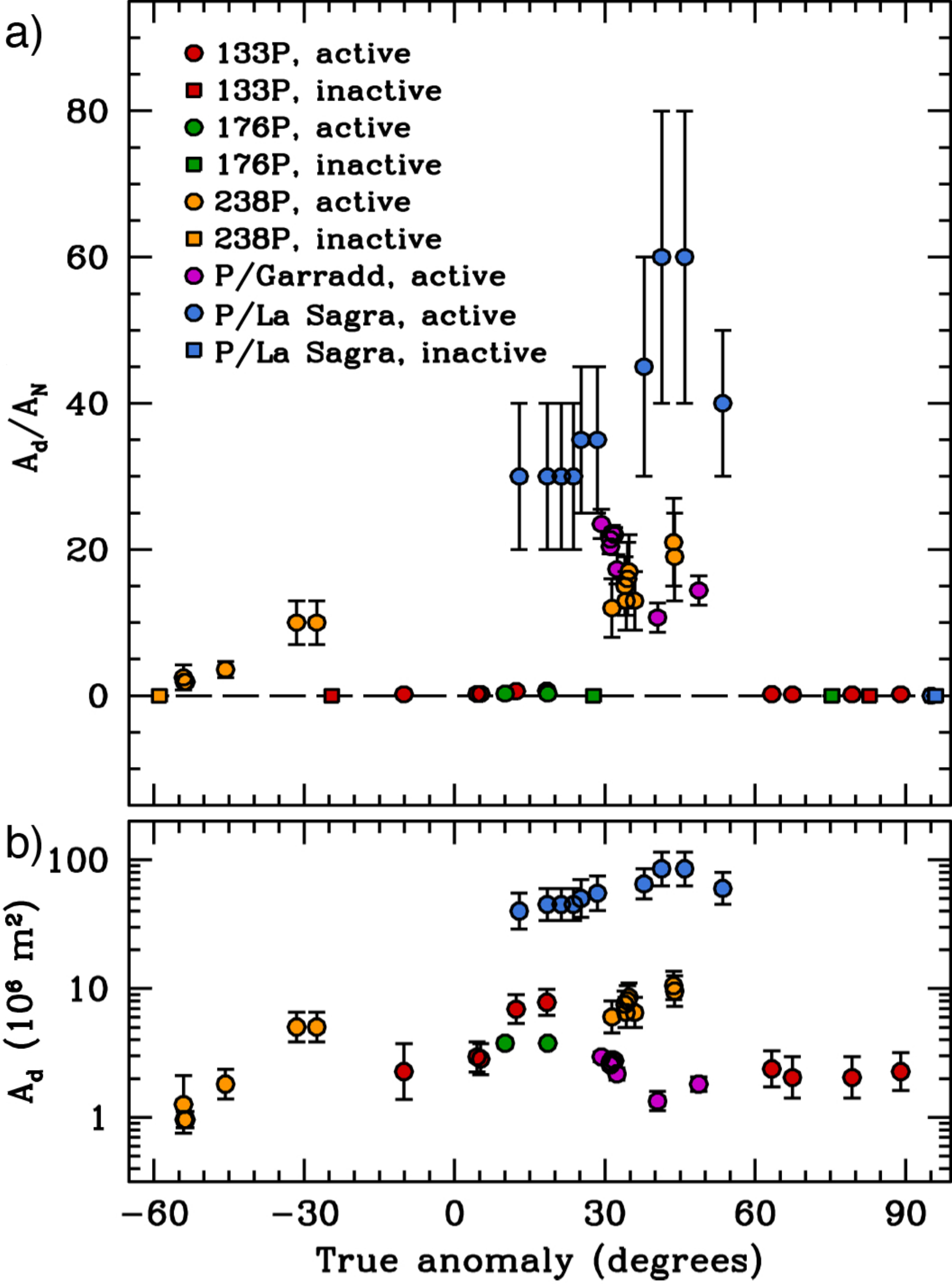}
\caption{Comparisons of (a) dust-to-nucleus ratios (by scattering cross-section) and (b) total dust scattering cross-sections, in $10^6$~km$^2$,
as functions of true anomaly measured for the five known MBCs:  133P \citep[red symbols;][]{hsi10b},
176P \citep[green symbols;][]{hsi11a}, 238P \citep[orange symbols;][]{hsi11b},
P/Garradd \citep[purple symbols;][]{jew09,mac12}, and P/La Sagra (blue symbols; this work).
Observations where activity was detected either visually or from photometry are marked with circular symbols, while
observations where no activity was detected are marked with squares.  For points where no error bars are visible, the amount of
uncertainty is equal to or less than the size of the plotted symbol.
}
\label{activity_comparison}
\end{figure}

To further understand P/La Sagra's place among the population of known MBCs, we plot $A_d/A_N$ for the first five known MBCs as functions of true anomaly, $\nu$ (Figure~\ref{activity_comparison}a).  We note a few interesting features of this plot.  First, we find quantitative confirmation that for their nucleus sizes, 238P and P/Garradd exhibit far more dust when active than 133P and 176P (also see Table~\ref{orbelems}).  This fact has of course always been qualitatively apparent. The observed morphologies of 238P and P/Garradd, when active, resemble those of highly active comets \citep{hsi09b,hsi11b,jew10}, whereas the nuclei of 133P and 176P dominate the surface brightness profiles of those objects, even at the peaks of their active periods.  Only comparatively faint dust tails and small amounts of photometrically-detected unresolved near-nucleus dust have attested to the cometary natures of these latter MBCs \citep{hsi04,hsi10b,hsi11a}.  As for P/La Sagra, we find that it exceeds even 238P and P/Garradd in the amount of dust it produces for the size of its nucleus, producing $\sim3$ times as much dust relative to its nucleus size at its peak as 238P and P/Garradd produced at their peaks, which in turn were $\sim50$ times as productive relative to their nucleus sizes as 133P and 176P.

We caution that given the different geocentric distances at which the observations in question were made, the near-nucleus photometry performed for each MBC measures dust contained within different physical distances from the nucleus.  As such, considering that other physical characteristics of the dust emission (e.g., grain sizes, bulk grain densities, and ejection velocities) for each MBC are unknown and likely also vary from object to object, these comparisons are best interpreted in terms of orders of magnitude.  We therefore find 238P, P/Garradd, and P/La Sagra to have similar levels of activity which are 1-2 orders of magnitude higher than those of 133P and 176P, implying similar ratios of fractional active surface areas for these objects.


To quantify the total amount of ejected dust exhibited by each comet, we also multiply the $A_d/A_N$ values for each comet by estimates of $A_N$ for each object (derived from nucleus radii listed in Table~\ref{orbelems}) to obtain absolute values for $A_d$.  We plot the results as functions of true anomaly in Figure~\ref{activity_comparison}b.  Interestingly, we find that despite the high $A_d/A_N$ values measured for 238P and P/Garradd, the total absolute amount of dust in those comets is approximately comparable to the amounts of dust present for 133P and 176P, which visually appear far less active in comparison.  P/La Sagra is seen to produce roughly an order of magnitude more dust in absolute terms than any of the other MBCs (assuming similar particle size distributions).

We therefore find the interesting result that within an order of magnitude, all five MBCs considered here exhibit similar total amounts of visible dust in their comae and tails, despite the wide variation in their observed morphologies.  Determining whether this finding is physically significant will require similar dust mass measurements for newly-discovered MBCs to investigate whether this trend is maintained as the sample of known objects grows.  We suggest that future research investigating the validity and implications of this possible trend may provide valuable insights into the origin and evolution of MBC activity.

\subsection{Spectroscopic Analysis\label{spectroscopy}}

\subsubsection{Data Reduction\label{specreduction}}

In the optical region of the electromagnetic spectrum, the emission band of the CN radical at 3880~${\rm \AA}$ is among the strongest among cometary gaseous species and therefore is considered to be the most sensitive probe for detecting gas resulting from sublimating ice. We search for evidence of the presence of the CN band in a one dimensional spectrum extracted from the spectral image using a $1\farcs0\times7\farcs5$ rectangular aperture centered on the continuum.  We use a relatively wide extraction aperture to incorporate as much light as possible from the surrounding coma. The sky background is reconstructed and subtracted using flanking regions extending from $9''$ to $15''$ from the nucleus.

\begin{figure}[ht]
\plotone{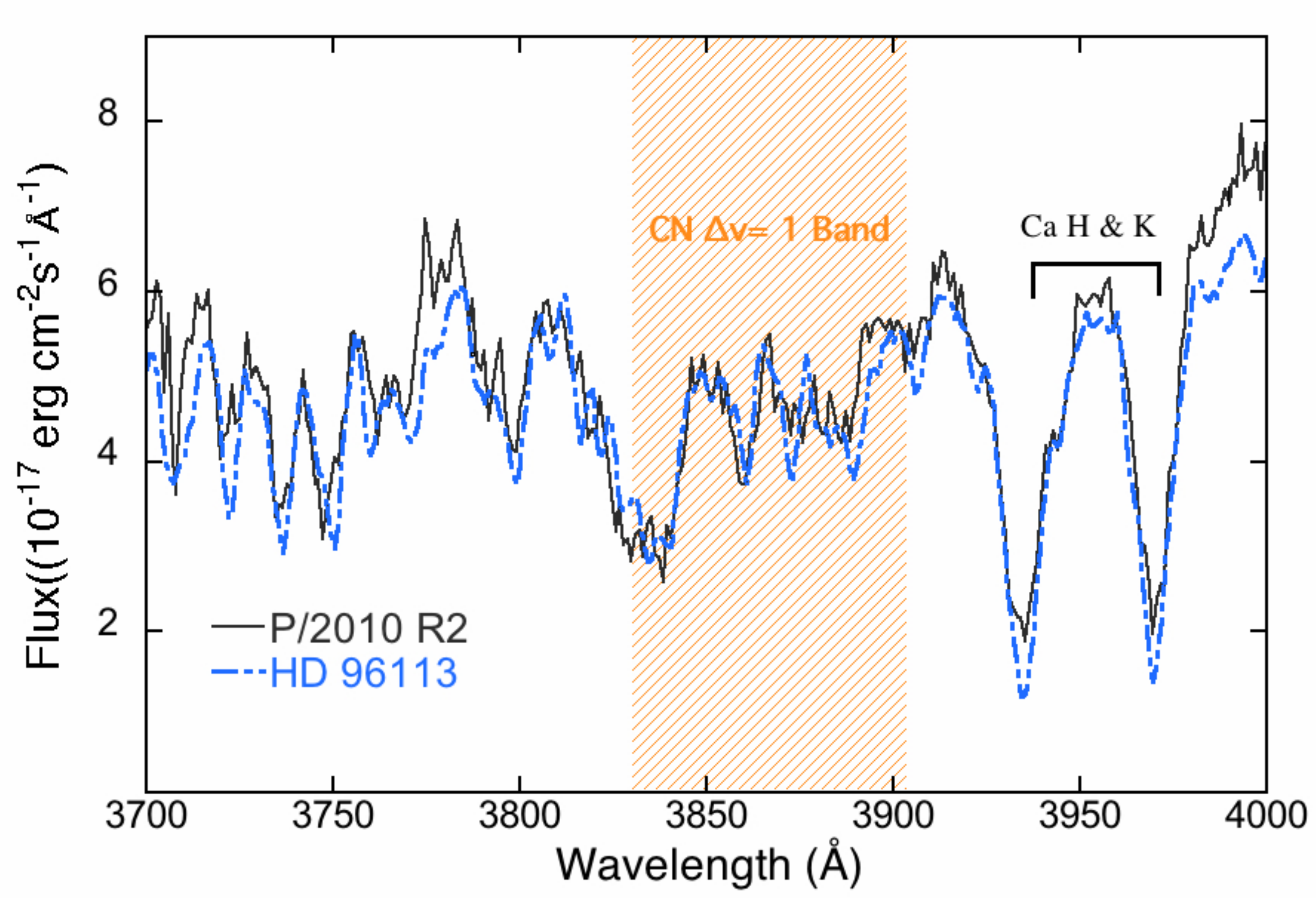}
\caption{Plots of the sky-background-subtracted and flux-calibrated spectrum of P/La Sagra (solid black line),
and the scaled spectrum of the solar analog, the G2V star HD 91163 (blue dashed line).
The shaded region indicates the wavelength region where the CN emission band is expected.
We note that the spectrum of P/La Sagra closely resembles that of HD 91163 and thus no emission features are observed
within the uncertainties of our measurements. }
\label{R2_f1}
\end{figure}

In Figure \ref{R2_f1}, we present the spectrum of P/La Sagra in the relevant spectral region from $\sim3700~{\rm \AA}$ to $\sim4000~{\rm \AA}$.  The continuum of the comet consists of scattered sunlight from dust particles in the inner coma and shows several absorption features in the Solar spectrum, such as the prominent Ca H and K absorption lines at 3933~${\rm \AA}$ and 3966~${\rm \AA}$ (Figure \ref{R2_f1}).  The shaded spectral region from 3830~${\rm \AA}$ to 3905~${\rm \AA}$ indicates where we expect to detect CN emission if any is present \citep{coc92}.

In order to correct for solar spectral features, the spectrum of a G2V solar analog star is needed. As discussed in Section~\ref{obs}, we were not able to observe a G2V star at the time of our P/La Sagra observations due to poor weather conditions. We instead adopt a flux-calibrated spectrum of G2V star HD 91163 which was obtained using the same instrument but was recorded a few months later on UT 2010 December 17 \citep{hsi12a}.  This spectrum was observed using the 600/4000 grism and thus has a higher spectral resolution than our P/La Sagra data, however, so the additional step of binning the HD~91163 spectrum to the same spectral resolution as our P/La Sagra spectrum is required. The scaled spectrum of HD 91163 is shown as the blue dashed line in Figure \ref{R2_f1}. We note that the spectra of P/La Sagra and HD 91163 are quite similar, particularly in the shaded region. However, minor discrepancies are observed at wavelengths $\lambda<3800~{\rm \AA}$ and $\lambda >3980~{\rm \AA}$.  These discrepancies may be due to the time and airmass differences between the comet observations and standard star observations. We find no evidence of CN emission in the shaded wavelength region.

\subsubsection{Upper Limits on Gas Production Rates\label{gaslimit}}

\begin{figure}[ht]
\plotone{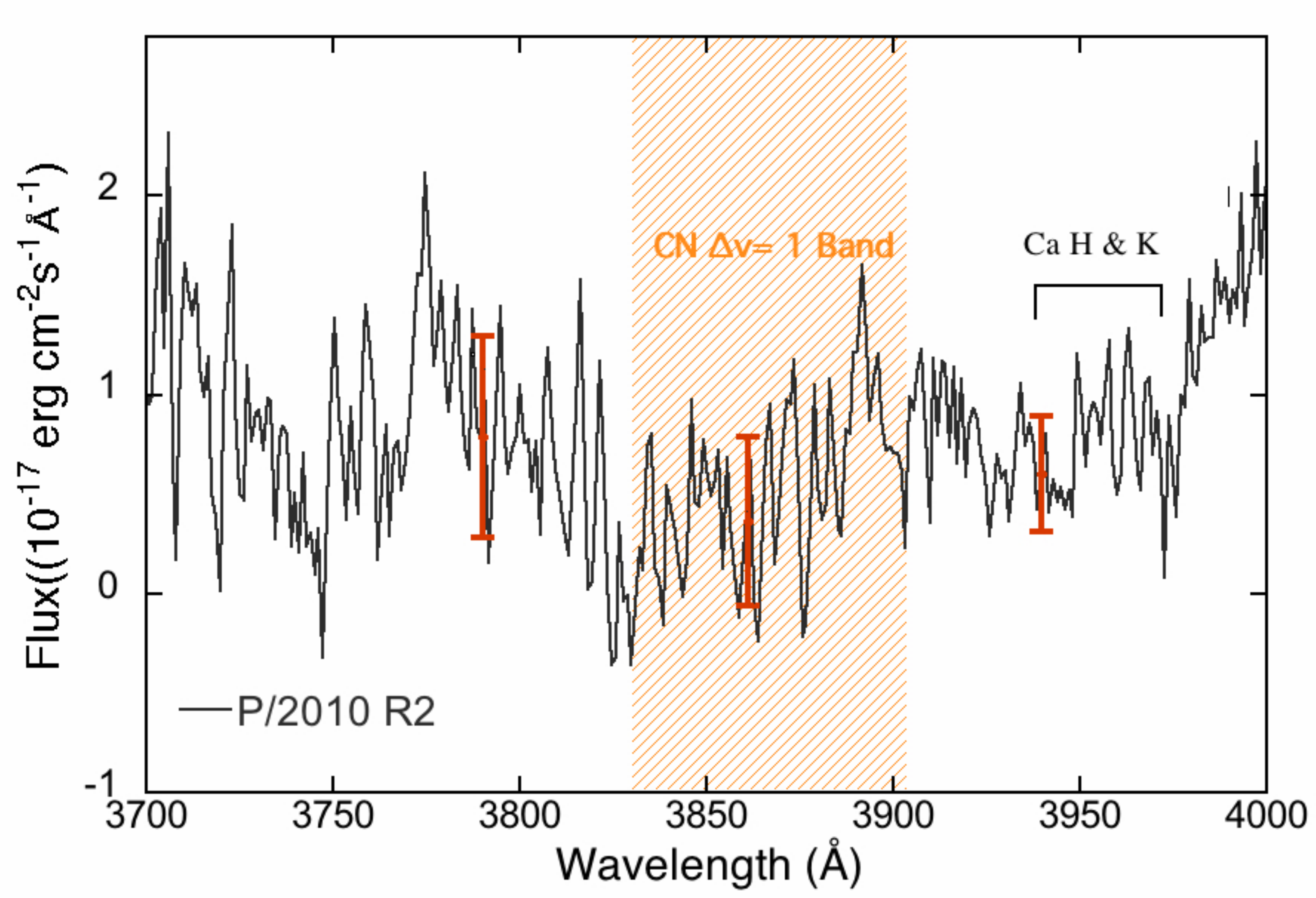}
\caption{The spectrum of P/La Sagra with the underlying solar continuum removed. The three red error bars show the 1-$\sigma$ uncertainties
in the three wavelength regions (3760-3830\AA, 3830-3900\AA, and 3900-3970\AA). }
\label{R2_f2}
\end{figure}

The continuum-removed spectrum of P/La Sagra is shown in Figure~\ref{R2_f2}. In principle, this spectrum shows emission from atoms and molecules in the inner coma. We measure the standard errors in the residual spectrum in three wavelength regions $70~{\rm\AA}$ in width: $3760~{\rm\AA}$ to $3830~{\rm\AA}$, $3830~{\rm\AA}$ to $3900~{\rm\AA}$, and $3900~{\rm\AA}$ to $3970~{\rm\AA}$. The standard errors on either side of the expected CN band are $5.1\times10^{-18}$ and $2.8\times10^{-18}$~erg~cm$^{-2}$~s$^{-1}$~${\rm \AA}^{-1}$, respectively, shown as red error bars in Figure \ref{R2_f2}. The standard error in the region of the expected CN band is $4.2\times10^{-18}$~erg~cm$^{-2}$~s$^{-1}$~${\rm \AA}^{-1}$. We choose the largest standard error of $5.1\times10^{-18}$~erg~cm$^{-2}$~s$^{-1}$~${\rm \AA}^{-1}$ as a conservative estimate of our observational uncertainty in the region of the expected CN emission band.  We therefore estimate a 3-$\sigma$ upper limit of $1.5\times10^{-17}$~erg~cm$^{-2}$~s$^{-1}$~${\rm \AA}^{-1}$. 

We calculate the integrated CN band flux by summing the estimated emission flux in the shaded CN region, obtaining $f_{\rm CN}=3.7\times10^{-16}$~erg~cm$^{-2}$~s$^{-1}$. We then convert the integrated flux to the total number of CN molecules using
\begin{equation}
L_{\rm CN} = 4\pi\Delta^2 f_{\rm CN}
\end{equation}
\begin{equation}
N_{\rm CN}= \frac{L_{\rm CN}}{g(R)}
\end{equation}
where $N_{\rm CN}$ is the number of CN molecules, $\Delta$ and $R$ are the geocentric and heliocentric distances, respectively, and $g(R)$ is the resonance fluorescence efficiency, which describes the number of photons scattered per second per radical for an optically thin coma, in erg~s$^{-1}$~molecule$^{-1}$ at a heliocentric distance of $R$. At the time of our observations, P/La Sagra had a radial velocity of $\dot{R}=1.1$~km~s$^{-1}$, for which $g(R=1~{\rm AU})=2.86\times10^{-13}$ \citep{sch10}. Substituting $f_{CN}=3.7\times10^{-16}$~erg~cm$^{-2}$~s$^{-1}$, $R=2.66$~AU, and $\Delta=1.82$~AU, we find an upper limit of $N=8.3\times10^{25}$ CN molecules in the extraction aperture.

A simple Haser model \citep{has57} was used to derive the CN production rate based on the number of CN molecules. The details of the model are described in \cite{hsi12a}. We find an upper limit to the CN production rate of $Q_{\rm CN}<6\times10^{23}$~mol~s$^{-1}$.  Taking average ratios measured for previously observed comets ($\log[Q_{\rm CN}/{Q_{\rm OH}}]=-2.5$; ${Q_{\rm OH}}/{Q_{\rm H_2O}}=90\%$) \citep{ahe95}, we estimate an upper limit on the water production rate of $Q_{\rm H_2O}<10^{26}$~mol~s$^{-1}$.  Incorporating the $Af\rho$ value calculated from data taken at the same time as these spectra (Section~\ref{activityevolution}; Table~\ref{photresults}), we find a lower-limit dust-to-gas ratio of $\log[Af\rho/Q_{\rm CN}]\gtrsim-22$, similar to values measured for other Jupiter-family comets \citep{ahe95}.  As this result is only an lower limit, however, it is not possible to ascertain from these data the degree to which P/La Sagra is depleted in volatiles relative to other classical comets from its apparently long residence time in the inner solar system (Section~\ref{stability}).

\subsubsection{Estimation of Effective Active Area}
We can estimate the effective active area of the nucleus of P/La Sagra by comparing the observed mass loss rate with theoretical rates obtained from applying a simple thermal model. To calculate the sublimation rate of surface ice, we solve the energy balance equation for the equilibrium temperature on the object's surface:
\begin{equation}
\frac{S_\odot(1-p)}{r^2}J(\theta) = \epsilon \sigma T_e^4+L\psi(T_e) 
\end{equation}
where $S_\odot=1360$~W~s$^{-1}$ is the solar constant, $p=0.05$ is the albedo, $R=2.66$~AU, $J(\theta)$ is a function of the solar zenith angle $\theta$ which describes the ratio between the absorbing and radiating area, $\epsilon=0.9$ is the emissivity, $\sigma$ is the Stefan-Boltzmann constant, $L=2.68\times10^6~{\rm J~{\rm kg^{-1}}}$ is the latent heat and $T_{e}$ is the effective surface temperature. The specific sublimation rate, $\psi$ (in kg m$^{-2}$ s$^{-1}$), of water is given by
\begin{equation}
\psi=P(T_e)\sqrt{\frac{m}{2\pi kT_e}}
\end{equation}
where $m$ is the mass of a water molecule, $k$ is Boltzmann's constant, and $P(T_e)$ is the saturated vapor pressure \citep{fan84}. 

In the fast rotator approximation (which yields the coldest surface temperature), the nucleus is considered isothermal and $J=\langle\cos\theta\rangle=1/4$. 
The hottest surface temperatures are derived in the sub-solar point (i.e., slow rotator) approximation, where the absorbing and radiating areas are identical and therefore $J=1$. We obtain sublimation rates of $1.3\times10^{-6}$~kg~m$^{-2}$~s$^{-1}$ and $6.1\times10^{-5}$~kg~m$^{-2}$~s$^{-1}$ for the cold and hot scenarios, respectively. Assuming an average mass loss rate of $\dot M_d\sim0.2$~kg~s$^{-1}$ (Section~\ref{activityevolution}), we estimate active areas of $1.5\times10^5$~km$^2$ and $3.3\times10^3$~km$^2$, corresponding to total active surface fractions of $\sim$3\% and $\sim$0.05\% (likely precise to an order of magnitude, at best), for the cold and hot cases, assuming dust-to-gas mass ratios of unity.  These derived active surface fractions are comparable to those of other Jupiter-family comets measured by \citet{ahe95}.
For comparison, both 133P and 176P are estimated to have active areas of $\sim10^2$ - $10^3$~m$^{2}$ \citep{hsi04,hsi09}, corresponding to active surface fractions of 0.0002 - 0.002\%, several orders of magnitude smaller than for P/La Sagra.

\subsection{Dynamical Analysis\label{dynamics}}

\subsubsection{Stability Analysis\label{stability}}

To better understand the likely origin of P/La Sagra, we also consider its dynamical properties.  \citet{jew09} raised the possibility that MBC P/Garradd could have originated in a different part of the asteroid belt from where we see it today.  As such, an important issue to consider is whether P/La Sagra is dynamically stable at its present location, or whether it is possible that it may have also originated elsewhere.

To assess P/La Sagra's dynamical stability, we randomly generate three sets of 100 test particles, each set with Gaussian distributions in orbital element space, centered on P/La Sagra's JPL-tabulated osculating orbital elements, where each set is characterized by a $\sigma$ value equal to 
the JPL-tabulated uncertainties (as of 2011 January 1) for each orbital element.  To investigate the stability of the region (in orbital element space) surrounding P/La Sagra, we also generate three additional sets of 100 Gaussian-distributed test particles each, where each set is again centered on P/La Sagra's osculating orbital elements but is characterized by a $\sigma$ value equal to $100\times$ the JPL-tabulated uncertainties.  We then use the N-body integration package, Mercury \citep{cha99}, to integrate the orbits of each set of test particles forward in time for 100 Myr (limitations on available computing resources unfortunately prevent us from conducting these simulations over significantly longer time periods, {\it e.g.}, 1-2~Gyr, in a reasonable amount of time), where we treat the eight major planets as massive particles and all test particles are considered to be massless.

\begin{figure}[ht]
\plotone{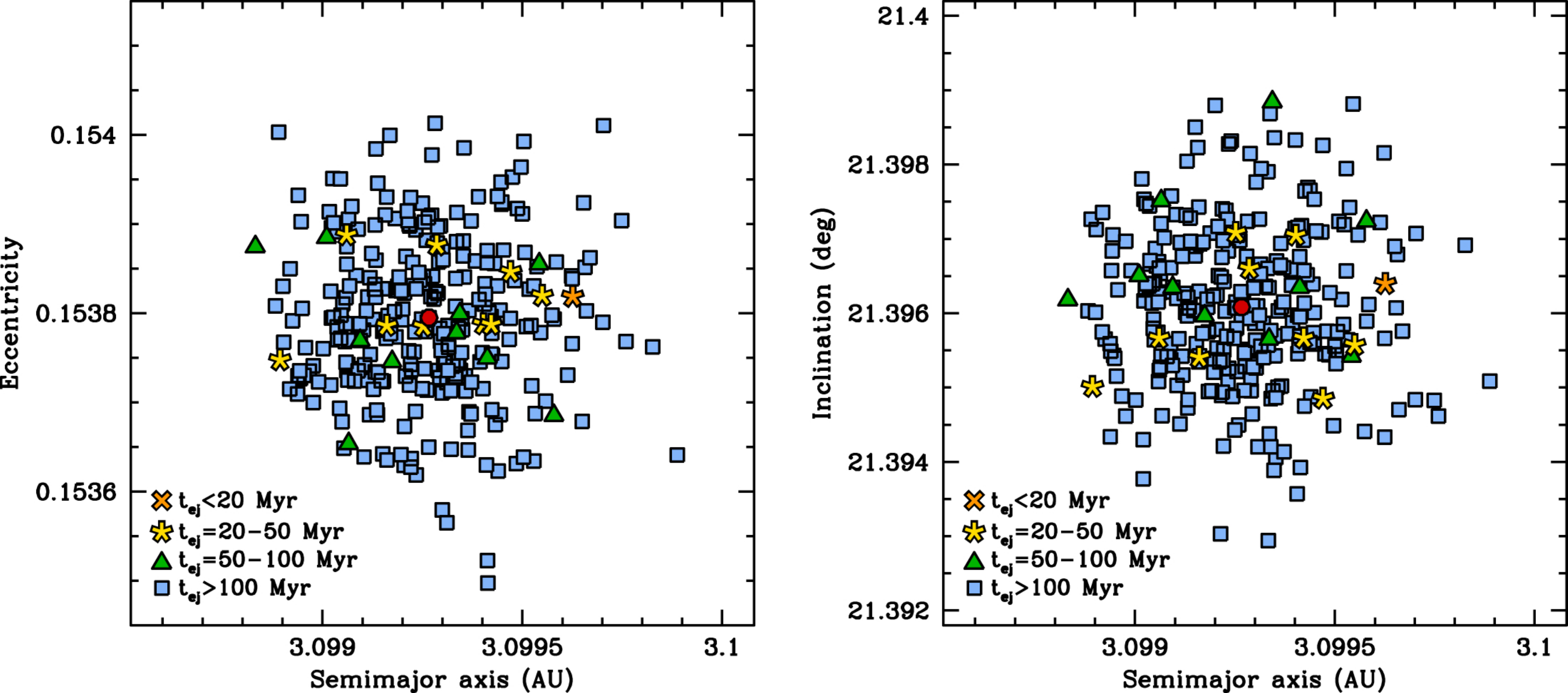}
\caption{Plots of semimajor axis versus eccentricity (left) and inclination (right) showing initial osculating elements of test particles
in three sets of 100 Gaussian-distributed particles each of which are subjected to a 100 Myr dynamical integration (Section~\ref{stability}).
Each set of test particles is centered on the current osculating orbital elements of P/La Sagra
(red circle) and characterized by a $\sigma$ value equal to the object's JPL-tabulated uncertainties.
Particles ejected in less than 20 Myr, between 20 Myr and 50 Myr, and between 50 Myr and 100 Myr are plotted with orange X symbols,
yellow asterisk symbols, and green triangles, respectively, while particles that are not ejected after the 100 Myr test period used
in our simulations are marked with blue squares.
}
\label{stability_1sig}
\end{figure}

\begin{figure}[ht]
\plotone{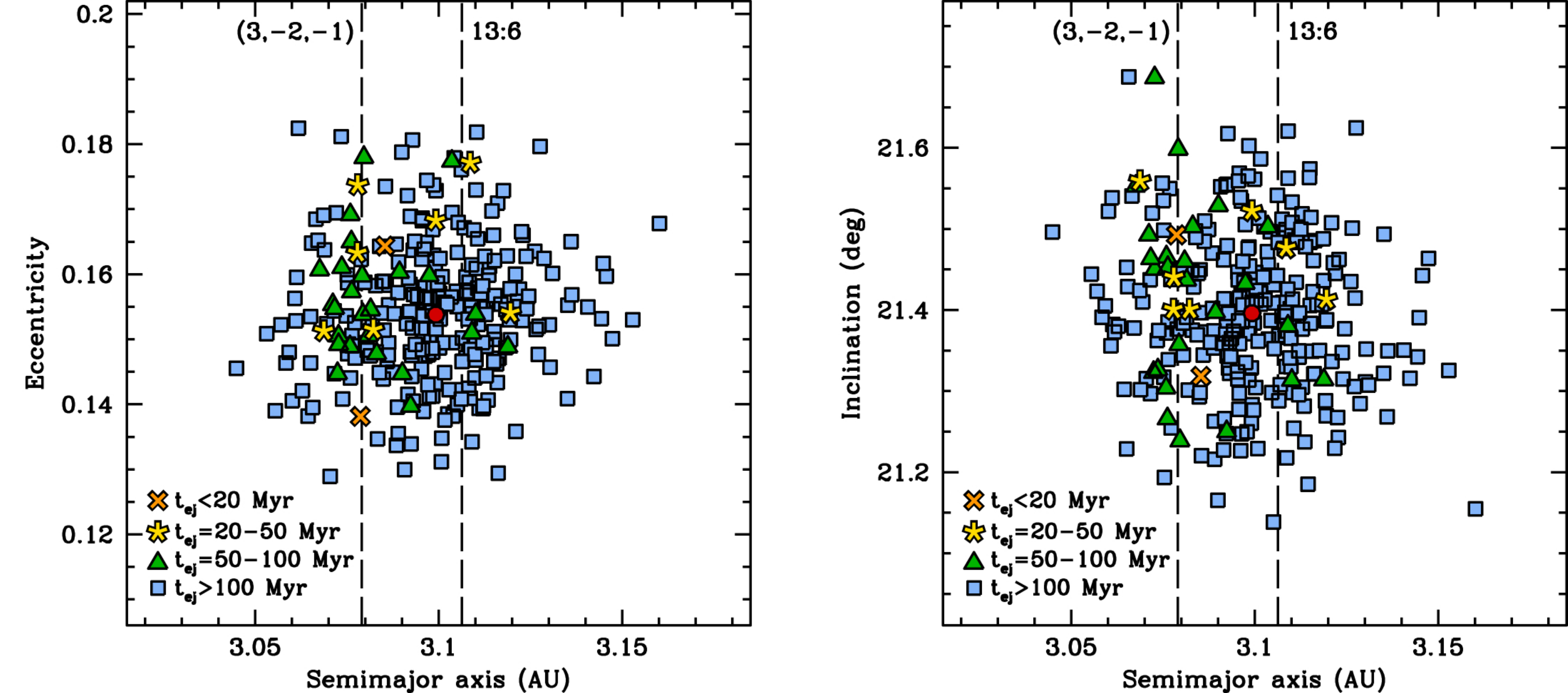}
\caption{Same as Figure~\ref{stability_1sig} for three sets of 100 Gaussian-distributed particles each, where each set is characterized
by a $\sigma$ value equal to 100$\times$ P/La Sagra's JPL-tabulated uncertainties.  The locations of the (3,-2,-1) three-body mean-motion
resonance (with
Jupiter and Saturn) at $a=3.0790$~AU and the 13:6 mean-motion resonance (with Jupiter) at $a=3.1063$~AU are plotted with dashed lines.
}
\label{stability_100sig}
\end{figure}

\begin{figure}[ht]
\plotone{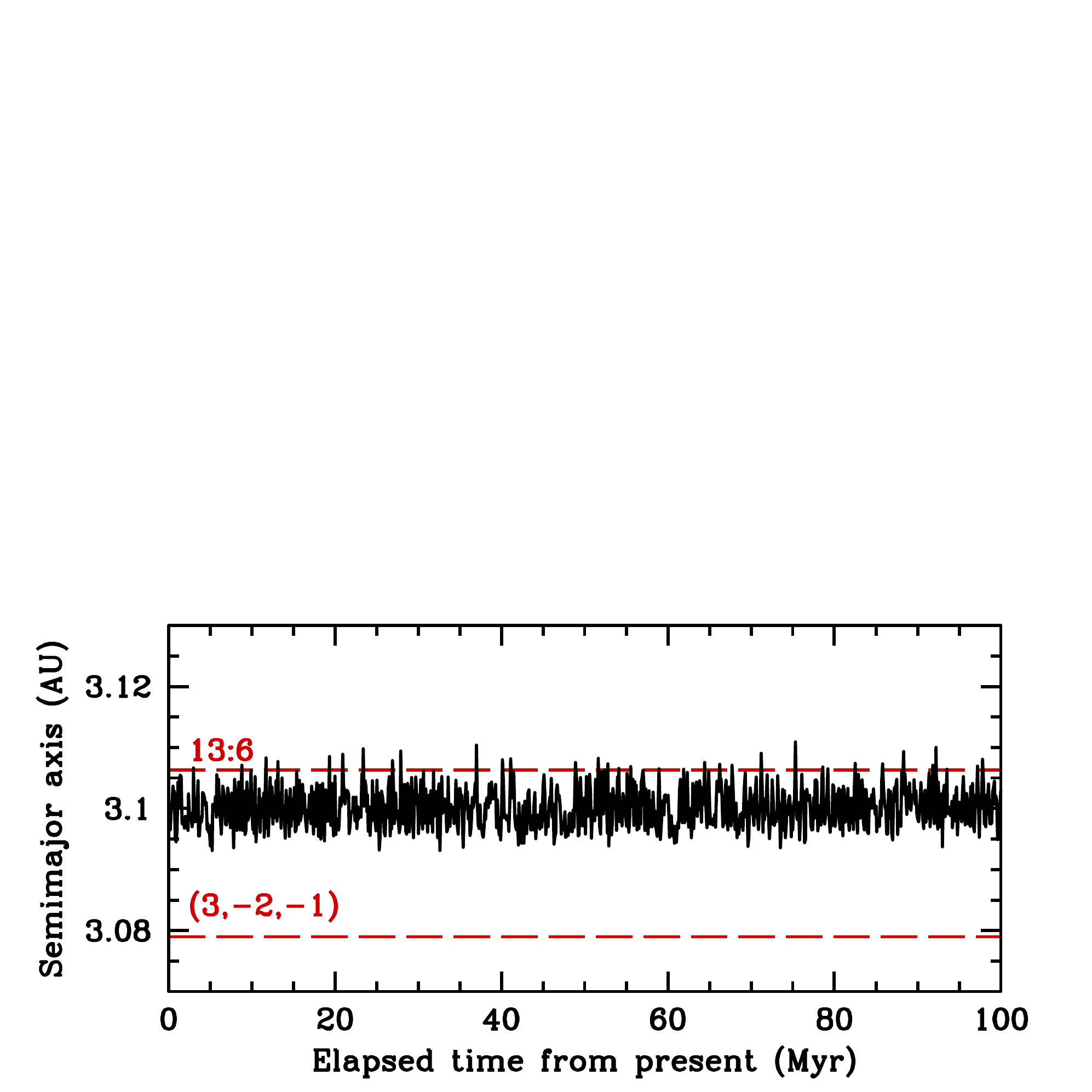}
\caption{Plot of P/La Sagra's semimajor axis (AU) as a function of time (Myr), marked by a black solid line, as simulated in the numerical
integrations described in Section~\ref{stability}, where the semimajor axes of the 13:6 mean-motion resonance with Jupiter and the (3,-2,-1) three-body
mean-motion resonance with Jupiter and Saturn are marked with red dashed lines.
}
\label{a_evolution}
\end{figure}

Performing these integrations for all six sets of test particles, we find that 7\% of the objects in the three 1-$\sigma$ test particle sets are ejected ({\it i.e.}, reach a heliocentric distance of $>50$~AU) from the asteroid belt over the course of 100~Myrs (Figure~\ref{stability_1sig}), and 11\% of the objects in the three 100-$\sigma$ test particle sets are ejected (Figure~\ref{stability_100sig}).  P/La Sagra itself is found to be stable over the 100~Myr test period, consistent with its Lyapunov time ($T_{\rm lyap}$), estimated by us to be $\sim$200~kyr  \citep[computed using the procedure described in][]{tsi03}, where a body is generally considered to be stable for $T_{\rm lyap}>100$~kyr.  The 1-$\sigma$ test particles that are ejected are not localized to particular regions of orbital element space near P/La Sagra, suggesting that their ejection may be a consequence of stochastic encounters with a weak mean-motion resonance (or multiple resonances), rather than the effect of a single dominant resonance.  Nearby resonances include the ``moderate-order'' 13:6 mean-motion resonance with Jupiter at $a=3.1063$~AU and the (3,-2,-1) three-body mean-motion resonance (with Jupiter and Saturn) at $a=3.0790$~AU \citep{nes98}.  Both resonances are more than 1-$\sigma$ away in $a$, given the uncertainties for P/La Sagra's osculating elements, but could conceivably affect the object as its osculating elements fluctuate over time (Figure~\ref{a_evolution}).

In our 100-$\sigma$ test particle simulations, the effect of the (3,-2,-1) three-body resonance is clear, with $\sim$70\% of the ejected test particles in those simulations having initial semimajor axes within 0.01~AU of the resonance.  The effect of the 13:6 resonance is comparatively less pronounced, with only $\sim$20\% of the ejected particles in these simulations starting within 0.01~AU of the resonance, despite the fact that this resonance is closer to P/La Sagra than the (3,-2,-1) resonance and therefore lies in a more highly populated region of our initial test particle distribution.  P/La Sagra itself has an initial osculating semimajor axis 0.06~AU of the 13:6 resonance, and even occasionally crosses the resonance during our simulations (Figure~\ref{a_evolution}), yet is found to be stable over the 100~Myr test period considered here, suggesting that this resonance only weakly affects objects in this region of orbital element space.


We conclude from this analysis that the region in orbital element space occupied by P/La Sagra is largely stable, implying that the object is likely to be native to its present-day location.  As such, its physical properties may be reflective of other objects in this region of the asteroid belt.  However, between the nearby 13:6 mean-motion resonance with Jupiter and (3,-2,-1) three-body mean-motion resonance with Jupiter and Saturn, enough dynamical instability is imparted to the region that a non-negligible fraction of simulated test particles escape in less than 100 Myr.  Due to the non-negligible instability of this region, we therefore cannot definitively conclude that P/La Sagra formed in situ (and did not dynamically evolve onto its present-day orbit from elsewhere in the main belt, or even elsewhere in the solar system), particularly over timescales longer than the 100~Myr test period considered in this analysis.  We further note that our analysis does not consider non-gravitational effects such as the Yarkovsky effect or the effects of asymmetric mass loss such as that associated with cometary jets.  As such, in future investigations of this object, particularly those which pertain to the comparison of its physical properties to those of other MBCs, the possibility that P/La Sagra may not be native to its current location cannot be discounted.

\subsubsection{Family Search\label{familysearch}}

Having investigated the dynamical stability of P/La Sagra and its surroundings, we seek to further characterize the object's dynamical environment by searching for evidence of an associated asteroid family, i.e., a grouping of dynamically linked asteroids likely to be the result of either the catastrophic fragmentation of a larger parent asteroid \citep{hir18}, or a significant cratering event on a still-existing asteroid \citep[{\it e.g.}, Vesta;][]{asp97}.  Such a search is particularly crucial for this object because of studies suggesting that MBCs may preferentially be found among families \citep{hsi04,hag09}, and possibly specifically among extremely young families \citep{nes08,hsi09}.  The latter hypothesis is based on the facts that MBC 133P appears to be a member of the $<10$-Myr-old Beagle family \citep{nes08}, and collisional devolatilization of near-surface ice of km-scale MBCs (like 133P and 176P) is expected to occur at rates incompatible with the existence of present-day activity, unless those MBCs are recently-produced fragments of larger asteroids with significant interior ice reservoirs \citep{hsi09}.

Due to P/La Sagra's high inclination, we cannot search for an associated dynamical asteroid family using analytically-determined proper orbital elements \citep{mil94}, as can be done for objects at lower inclinations.
The complication stems from the fact that analytically-computed proper elements are significantly less accurate for objects with high eccentricities or inclinations.  The development of numerical methods for computing so-called synthetic proper elements for characterizing the long-term behavior of high-inclination and high-eccentricity objects resolved this problem \citep{kne00}, an advance that was employed by \citet{gil06} and \citet{nov11} to conduct systematic searches for families among high-inclination asteroids.

\begin{figure}[ht]
\plotone{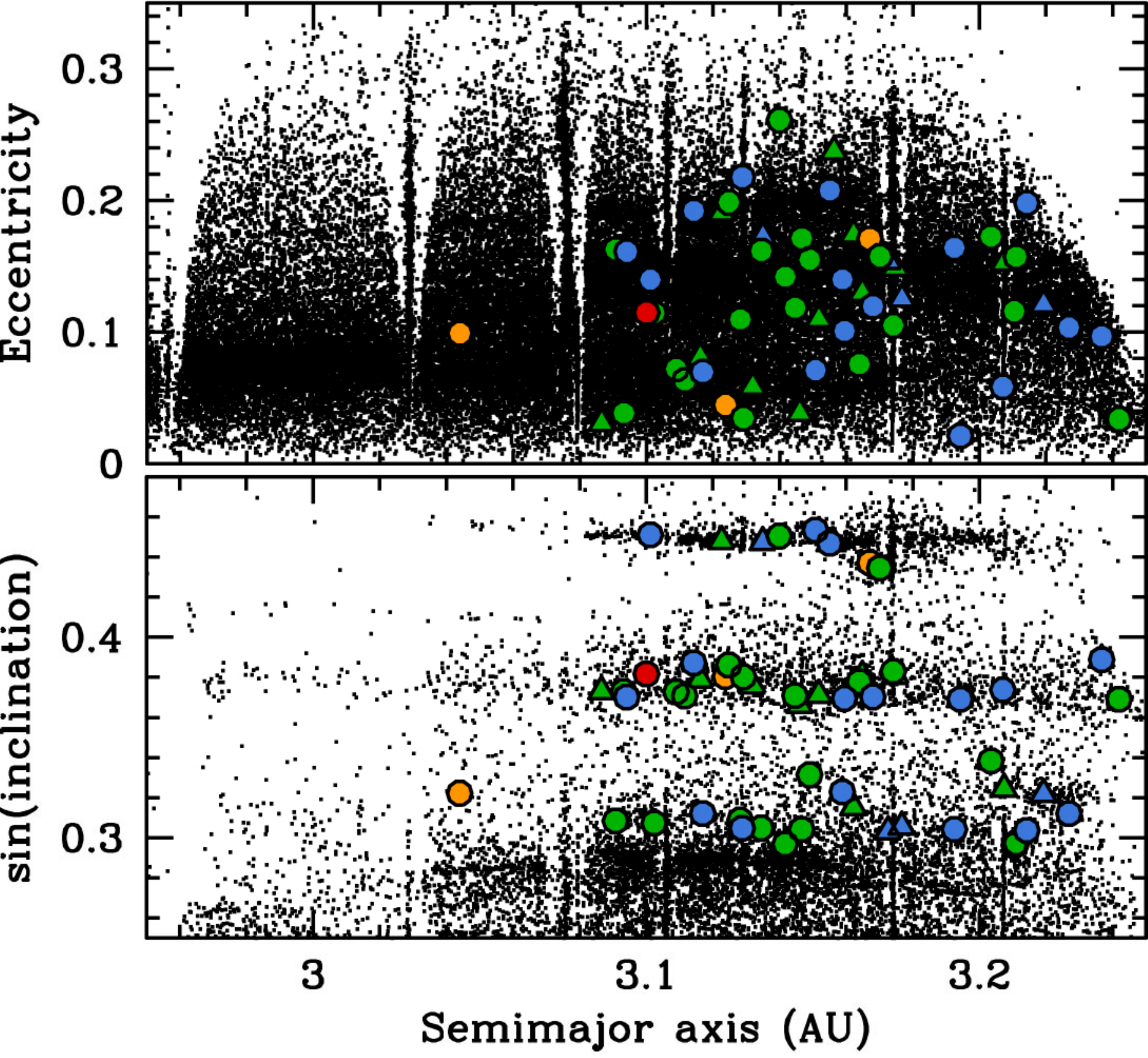}
\caption{Synthetic orbital element plots of eccentricity (upper panel) and sine of inclination (lower panel) versus
semimajor axis of outer main-belt asteroids (small black dots), where the central bodies of families and clumps
identified by \citet{gil06} are marked with blue and green triangles, respectively, and central bodies of families, clumps,
and clusters identified by \citet{nov11} are marked with blue, green, and orange circles, respectively.
The location of P/La Sagra in synthetic proper element space is plotted with a red circle.
}
\label{aeaifamilies}
\end{figure}

For our analysis of P/La Sagra, we first compute the object's own synthetic proper elements --- proper semi-major axis ($a_p$), proper eccentricity ($e_p$), and proper inclination ($i_p$) --- using the procedure described in \citet{kne00} and osculating orbital elements from the JPL Small-Body Database Browser.  We list these results in Table~\ref{lasagcluster}.  We then plot these newly computed synthetic proper elements in $a_p-e_p$ and $a_p-i_p$ space along with the synthetic elements of all of the families (as well as smaller or less-significant groupings designated as clumps or clusters) identified by \citet{gil06} and \citet{nov11} to check for plausible close associations of P/La Sagra with any of these previously identified groupings (Figure~\ref{aeaifamilies}).  We find no such associations.

\begin{figure}[ht]
\centering
\plotone{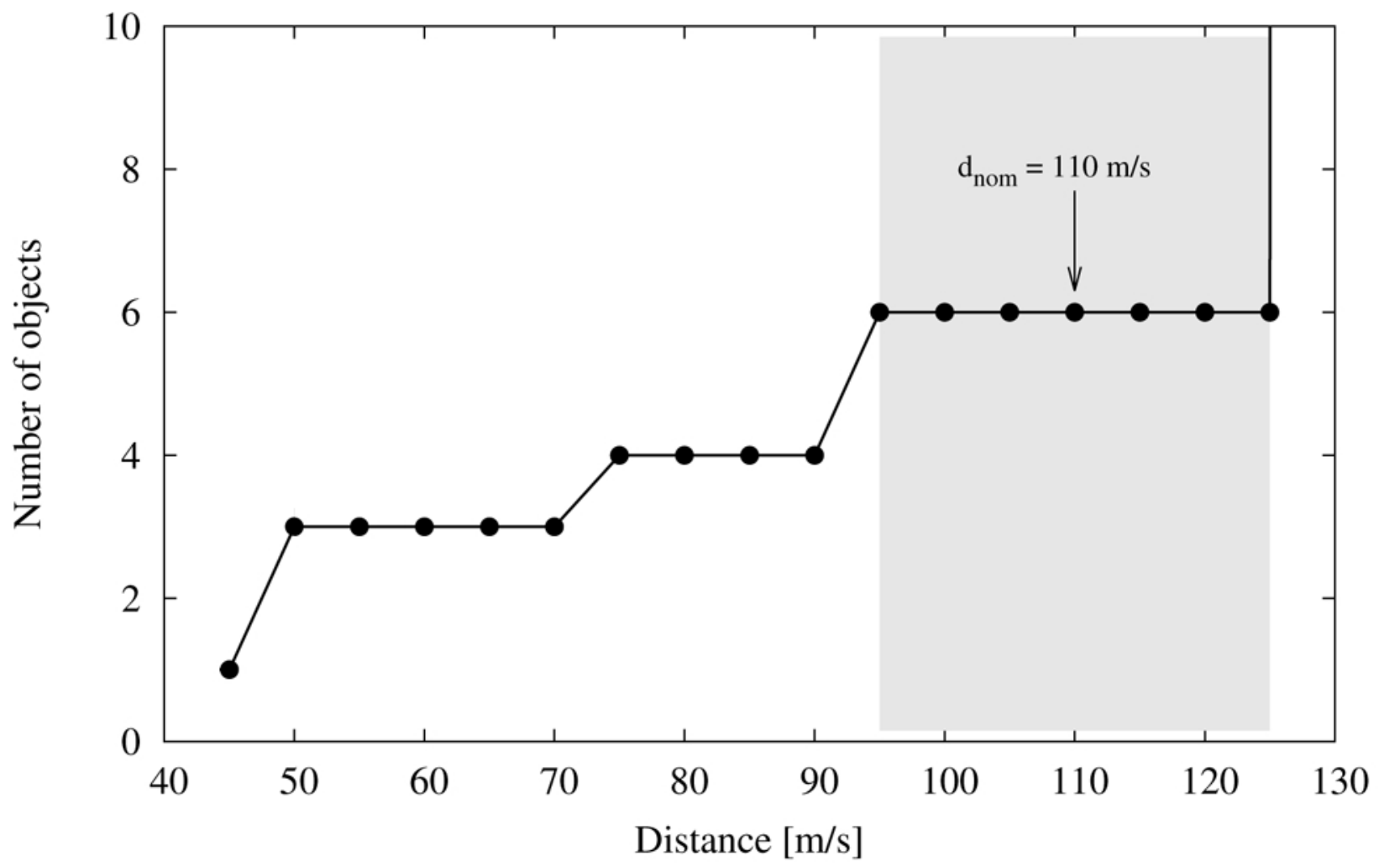}
\caption[]{Number of objects associated with the P/La Sagra cluster.
The six members determined to belong to this cluster are all linked within 95 m~s$^{-1}$,
with no more linkages found to other asteroids until 125 m~s$^{-1}$
when the group merges with the local background population.  
Our choice for the nominal cutoff distance for the cluster of 110 m~s$^{-1}$ is marked.}
\label{nfv}
\end{figure}

Finding no link between P/La Sagra and any currently known asteroid families, we turn to our own search for dynamically associated asteroids to determine whether a clustering specific to P/La Sagra may have missed simply because this newly discovered object was not considered in previous family searches.  To do so, we add our computed proper elements for P/La Sagra to an updated list of synthetic proper elements for high-inclination asteroids \citep{nov11}, and then employ the Hierarchical Clustering Method \citep[HCM;][]{zap90,zap94} to search for dynamically associated asteroids.  Results from this analysis are plotted in Figure~\ref{nfv} where we show the number of asteroids associated with P/La Sagra as a function of cutoff ``distance'' (which, despite its name, has units of velocity).

We find just six objects in P/La Sagra's dynamical vicinity (including P/La Sagra itself) within a cutoff distance of 120~m~s$^{-1}$ (Table~\ref{lasagcluster}).  Confirming our search (described above) for associations of P/La Sagra with previously identified groupings, none of these objects belong to any of the families, clumps, or clusters found by \citet{gil06} or \citet{nov11}.  According to the criteria used by Novakovi\'c et al.\ for the outer belt region, this group of objects is neither compact nor large enough to be classified as a family or clump.  At best, it can be classified as a cluster, defined by \citet{nov11} as a compact group of objects that could share a common physical origin, but does not formally satisfy criteria for being a family or clump, usually because the number of members that it contains is too small.

To obtain a quantitative assessment of the likelihood that members of the P/La Sagra group share a common physical origin, we use a method employed by \citet{nes06} to estimate the probability that a certain cluster of objects is the result of random chance.  Confining our analysis to the high-inclination region of the outer main belt ($2.825<a_{p}<3.278$ AU, $e_{p}<0.4$, $0.3<\sin(i_{p})<0.55$), we generate 10,000 random distributions of 9569 synthetic objects (the same number of objects currently known to populate this region) in proper orbital element space.  We then perform HCM analysis on these synthetic populations to search for possible groupings with at least 6 objects within cutoff distances of 95~m~s$^{-1}$.  The results of this analysis show that there is a significant probability ($\sim30$\%) that the P/La Sagra cluster could in fact be due to chance.  While these results still leave a $\sim70$\% chance that this grouping is not random and therefore could have a physical origin, more information is certainly needed to establish whether or not this clustering can be considered real.

We compute Lyapunov times for members of the P/La Sagra group (Table~\ref{lasagcluster}), finding the orbits of all six bodies to be stable ($T_{\rm lyap} >$100 kyr). This fact provides an opportunity to apply the backward integration method \citep[BIM;][]{nes02} to this group. BIM is a means for estimating the age of a $<$20-Myr-old family that is based on the premise that immediately after the disruption of a parent body, the orbits of the resulting fragments must have been nearly the same,
and has been successfully applied to estimate the ages of the Karin cluster and Veritas family 
\citep{nes03}.

In the case of the group of asteroids surrounding P/La Sagra, the purpose of BIM analysis is twofold.  If a suspicious clustering in the longitude of the ascending node ($\Omega$) or the longitude of perihelion ($\varpi$) can be found, this analysis would not only allow us to estimate the age of the group, but would also provide support for the hypothesis that these objects share a physical origin in the first place.  In the outer belt, however, application of this method is generally complicated by large gradients of secular frequencies with $a_{p}$, particularly when considering $\varpi$ in the vicinity of the 2:1 resonance with Jupiter \citep{nes03}.  Although the P/La Sagra group is not particularly close to this resonance, the gradient $dg/da_{p}$, where $g$ is the frequency of the longitude of perihelion, is still relatively large in this region ($\sim$0.1\degr~yr$^{-1}$~AU$^{-1}$).  As such, for the purpose of our study we only consider $\Omega$.

\begin{figure}[ht]
\centering
\plotone{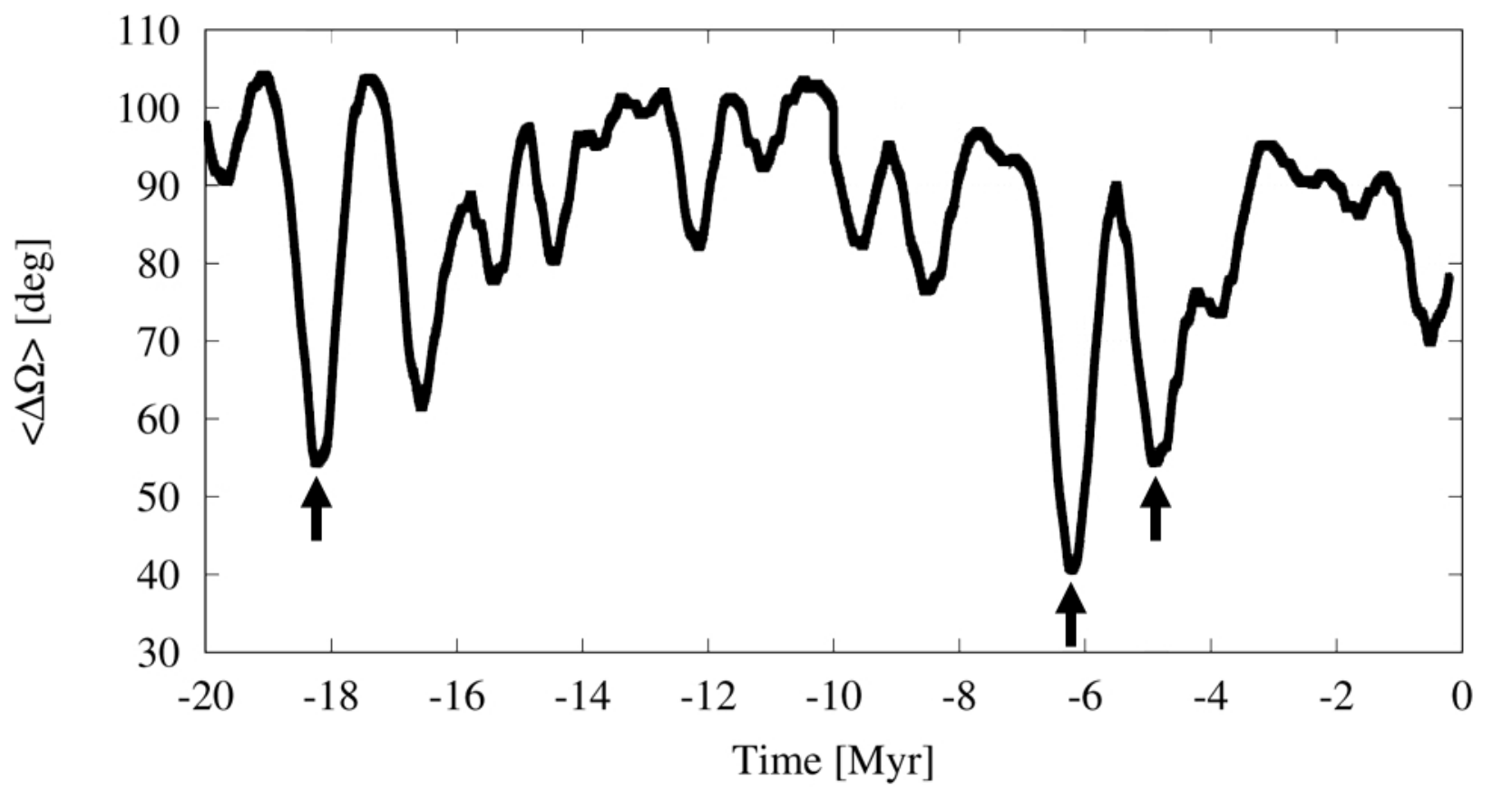}
\caption{Average differences $\langle \Delta \Omega \rangle$ of the longitudes of the ascending nodes
for members of the P/La Sagra cluster. 
Three significant clusterings at 4.9, 6.2 and 18.2 Myr in the past are marked with arrows.}
\label{clustering}
\end{figure}

To investigate whether or not a clustering in $\Omega$ exists, we numerically integrate the orbits of all six known members of the P/La Sagra group for 20 Myr backward in time.  We then calculate the average difference, $\langle \Delta \Omega \rangle$, calculated for all possible combinations of objects, where $\Delta \Omega=\Omega_{i} - \Omega_{j}$, and $i$ and $j$ denote two bodies under consideration.  Plotting the results of this analysis (Figure~\ref{clustering}), we find three clusterings within 60 degrees. The clusterings at about 4.9 and 18.2~Myr ago are less significant than the one occurring about 6.2~Myr ago. This implies that the age of the cluster is likely close to 6.2 Myr, but due to the small number of asteroids involved, it is not possible to draw a definitive conclusion about the physical significance of this cluster.

The results shown in Figure~\ref{clustering} are obtained in a purely gravitational model.  Including other forces ({\it e.g.}, the Yarkovsky effect) could affect the positions and depths of various minima \citep[{\it cf}.][]{nov10}.  We believe, however, that such an analysis should be deferred pending the discovery of more members of the cluster so that more statistically meaningful results can be obtained than are currently achievable with just six possible cluster members.
In the meantime, the determination of colors or acquisition of spectra of the currently suspected cluster members could provide some preliminary insights into the likelihood that these objects are physically related.

\section{DISCUSSION\label{discussion}}

To date, the only published analysis of P/La Sagra prior to this work was presented by \citet{mor11},
who focused on numerical modeling of object's dust emission.  They found that the activity
can be plausibly modeled by an episode of continuous dust emission, consistent with a sublimation-driven
dust ejection process and inconsistent with impulsive dust ejection as would be expected from an impact,
beginning shortly before the object's perihelion passage and persisting over $>$7~months.
Furthermore, they found that the dust cloud's morphology was best reproduced by anistropic
dust emission and indicated that their modeling suggests that P/La Sagra's activity
could be seasonally modulated in a manner similar to that of other MBCs \citep[{\it e.g.},][]{hsi04,hsi11a}.
In our photometric analysis of the dust cloud (Section~\ref{activityevolution}), we
found steadily increasing dust mass between August 2011 and December 2011, indicative of ongoing dust
production, and consistent with the conclusion by \citet{mor11} that P/La Sagra's dust emission appears to have
persisted over an extended period of time, and therefore was likely to be cometary ({\it i.e.}, sublimation-driven)
in nature.

While the conclusion that P/La Sagra's dust emission is driven by the sublimation of volatile ices appears
to be at odds with the results of our spectroscopic analysis (Section~\ref{spectroscopy}), we note that the
extreme difficulty of detecting sublimation products in a cometary coma at the distance of the main belt means
that the non-detection of gas emission in such observations cannot be considered conclusive evidence against
the presence of gas \citep[{\it cf}.][]{hsi12a}.  Rather, such non-detections should only be interpreted as
evidence that the levels of gas emission for all MBCs for which spectroscopy has
been obtained to date \citep[133P, P/Garradd, and P/La Sagra;][and this work]{lic11,jew09} have simply been
below the sensitivity levels of those observations.  However, the aforementioned observations were all obtained using
large ground-based telescopes (the European Southern Observatory's 8~m Very Large Telescope
in Chile, and the 10~m Keck I telescope in Hawaii), and so future attempts to detect gas in MBCs
may need to focus on significantly more active MBCs ({\it i.e.}, with higher gas emission levels) since significantly more powerful
observing facilities are not currently available.  It may also be
possible that given the large distances and low activity levels (relative to other comets observed at much
smaller heliocentric and geocentric distances) of the MBCs, conclusive evidence of gas emission may only ever
be obtainable from a close-approaching spacecraft equipped to make such measurements.

Despite the lack of a successful gas detection (Section~\ref{spectroscopy}), we conclude from the numerical modeling of \citet{mor11}
and the morphological and photometric analyses presented in this work
(Sections~\ref{imageanalysis} and \ref{photometry}) that P/La Sagra's activity is likely to be
sublimation-driven.  As such, the object is likely to be a bona fide MBC like 133P, 176P, 238P, and P/Garradd,
where dust is ejected via gas drag from sublimating ice, and not as ejecta from a recent impact as is suspected
in the cases of disrupted asteroids, P/2010 A2 and (596) Scheila \citep[{\it e.g.},][]{jew10,jew11,sno10,bod11}.

Definitive criteria for identifying the true nature of comet-like activity remain elusive, however \citep[{\it cf}.][]{hsi12a}.
In the case of P/La Sagra, \citet{mor11} show that comet-like dust emission can plausibly explain the appearance
of P/La Sagra's dust tail, but by no means show that their solution is unique, a common weakness in virtually
all dust modeling efforts where models are underconstrained by available observations.
Given the case of P/2010 A2's long-lived dust tail \citep{jew10,sno10}, the longevity of P/La Sagra's activity is likewise not
incontrovertible evidence of sublimation-driven dust emission.  Even the apparently steadily increasing dust mass in P/La Sagra's coma
and tail (Section~\ref{activityevolution}) could be considered inconclusive evidence of sublimation
given the potential that fragmenting dust particles ejected in an impulsive event could mimic ongoing dust
production as in the case of the 2007 outburst of comet 17P/Holmes \citep{hsi10a}.  

At this time, recurrent activity, separated by intervening periods of inactivity, appears to be the only reliably observable indicator of sublimation-driven dust emission that cannot also be plausibly explained by a scenario involving impact-driven dust ejection \citep{hsi11c}.  Repeated impacts on individual asteroids are unexpected from either theoretical calculations or empirical observations.  The tendency of episodes of repeated activity to occur over similar arcs of each object's orbit makes such behavior even more difficult to explain as the action of random impacts rather than the effect of periodic cometary sublimation.  Two of the six currently known MBCs have been observed to exhibit recurrent activity to date \citep[133P and 238P;][]{hsi04,hsi10b,hsi11b}, while 176P is expected to exhibit renewed activity shortly \citep{hsi11a}.  Confirmation of renewed activity in 176P will be crucial for supporting the hypothesis that its activity is cometary in nature.  The remaining three MBCs (P/Garradd, P/La Sagra, and P/2006 VW$_{139}$), on the other hand, were actually discovered so recently that they have not yet completed full orbits since being discovered, but as they do (P/Garradd in early 2013, P/La Sagra in mid-2015, and P/2006 VW$_{139}$ in mid-2016), observations to search for recurrent activity are highly encouraged.

\section{SUMMARY \& CONCLUSIONS}

We present a wide-ranging study of the recently discovered comet P/2010 R2 (La Sagra) including photometric and spectroscopic observations and dynamical analyses of the object.  Key results are as follows:
\begin{enumerate}
\item{A year-long observational monitoring campaign of P/La Sagra from August 2010 to August 2011 show that the morphology of its dust cloud undergoes substantial evolution, exhibiting a strong coma and a single tail from August 2010 to November 2010, and a coma and two tails from December 2010 to February 2011, before diminishing to a faint, detached dust trail by August 2011.  A long-lived antisolar dust tail and the steadily increasing scattering surface area measured for the dust cloud over much of this period suggests that dust production was ongoing over a period of several months, behavior consistent with dust ejection via the sublimation of volatile ice and inconsistent with dust ejection via the impulsive action of an impact on the surface of P/La Sagra.
  }
\item{We estimate that the nucleus of P/La Sagra has an absolute magnitude of $H_R=17.9\pm0.3$~mag, corresponding to an approximate effective radius of $r_N=0.7\pm0.1$~km.  The $B-R$ color of the nucleus ($B-R=0.9\pm0.1$~mag) is approximately solar, as are the $B-R$ colors of the active comet measured on two separate occasions in 2010.
  }
\item{Optical spectroscopic observations of the active P/La Sagra 20 days after it was discovered reveal no evidence of CN emission.  Based on the sensitivity of our observations, we find upper-limit CN and H$_2$O production rates of $Q_{\rm CN} < 6\times10^{23}$~mol~s$^{-1}$ and $Q_{\rm H_2O} < 10^{26}$~mol~s$^{-1}$, respectively.
  }
\item{Numerical integrations show that P/La Sagra is largely dynamically stable, indicating it is likely native to its current location in the main belt.  The surrounding region in orbital element space, however, is moderately unstable with two resonances, the 13:6 mean-motion resonance with Jupiter and the (3,-2,-1) three-body mean-motion resonance with Jupiter and Saturn, present nearby.
  }
\item{We find that P/La Sagra is not associated with any known asteroid families, though find a small cluster of asteroids in its dynamical vicinity. Due to the small number of objects in this cluster at the current time, however, we cannot confirm whether members of this cluster in fact resulted from the recent fragmentation of a larger parent asteroid, or if they are simply clustered in orbital element space by chance.
  }
\end{enumerate}
Despite our unsuccessful attempt to detect gas emission from P/La Sagra, we conclude from the long-lasting and steadily evolving morphology of the object's dust cloud that its activity is likely to be sublimation-driven, and unlikely to be the result of an impulsive impact.  As such, it is likely to be a genuine comet, making it just the fifth MBC discovered (out of a total of six known to date).  Numerical simulations indicating that P/La Sagra is likely to be native to its current location in the main belt mean that its physical composition is likely to be representative of other objects in the same region.  A search for other MBCs in this region of orbital element space could prove fruitful.  It has been hypothesized, however, that present-day MBC activity may require triggering by small impacts that excavate near-surface ice and expose it to the Sun \citep{hsi04,hsi09}.  As such, we caution that the lower rate of potential activity-triggering impacts at high inclinations \citep{far92} means that, overall, the rate of currently active MBCs in this region is likely to be lower than at lower inclinations.

In any event, the presence of P/La Sagra in a completely disparate region of orbital element space from the other MBCs underscores the caveat by \citet{hsi09} that the currently known population is far too small to make any conclusions about the abundance and distribution of such objects in the asteroid belt.  Current and upcoming systematic searches of the entire asteroid belt for MBCs, now starting to see success in the case of PS1 \citep{hsi11c,hsi12b}, will be critical for ascertaining the true abundance and distribution of icy objects in the inner solar system.

\begin{acknowledgements}
Support for this work was provided by NASA to H.H.H. through Hubble Fellowship grant HF-51274.01 awarded by the Space Telescope Science Institute, which is operated by the Association of Universities for Research in Astronomy, Inc., for NASA, under contract NAS 5-26555.
B.Y. and N.H. acknowledge support through the NASA Astrobiology Institute under Cooperative Agreement No.\ NNA08DA77A issued through the Office of Space Science.
B.N. is supported by the Ministry of Education and Science of Serbia, under Project 176011, while
M.G. is funded by the Academy of Finland grant \#137853.
C.S. additionally acknowledges funding from the European Union Seventh Framework Programme (FP7/2007-2013) under grant agreement no. 268421.
We are grateful to John Dvorak and Richard Morriarty, Greg Wirth, Heather Hershley, Scott Dahm, Marc Kassis, Luca Rizzi, and Julie Renaud-Kim, and Jay Rhee, Jonathan Kemp, and Chad Trujillo for their assistance in obtaining observations on the UH~2.2~m, Keck, and Gemini telescopes, respectively, and Ilona Busenbender and Eduardo Solares for technical assistance at the INT.
We also thank Alberto Cellino for valuable discussions about this work.
Acquisition of some of the data presented was enabled using the PS1 System operated by the PS1 Science Consortium (PS1SC) and its member institutions.  The PS1 Survey has been made possible through contributions of the Institute for Astronomy, the University of Hawaii, the Pan-STARRS Project Office, the Max-Planck Society and its participating institutes, the Max Planck Institute for Astronomy, Heidelberg and the Max Planck Institute for Extraterrestrial Physics, Garching, The Johns Hopkins University, Durham University, the University of Edinburgh, Queen's University Belfast, the Harvard-Smithsonian Center for Astrophysics, and the Las Cumbres Observatory Global Telescope Network, Incorporated, the National Central University of Taiwan, and the National Aeronautics and Space Administration under Grant No.\ NNX08AR22G issued through the Planetary Science Division of the NASA Science Mission Directorate.
The Gemini Observatory is operated by the Association of Universities for Research in Astronomy, Inc., under a cooperative agreement with the NSF on behalf of the Gemini partnership: the National Science Foundation (United States), the Science and Technology Facilities Council (United Kingdom), the National Research Council (Canada), CONICYT (Chile), the Australian Research Council (Australia), Minist\'{e}rio da Ci\^{e}ncia e Tecnologia (Brazil) and Ministerio de Ciencia, Tecnolog\'{i}a e Innovaci\'{o}n Productiva (Argentina).
Observations at the Danish 1.54m telescope were supported by the Danish Natural Science Research Council (FNU).
Funding for SDSS-III has been provided by the Alfred P. Sloan Foundation, the Participating Institutions, the National Science Foundation, and the U.S. Department of Energy Office of Science. The SDSS-III web site is {\tt http://www.sdss3.org/}.  SDSS-III is managed by the Astrophysical Research Consortium for the Participating Institutions of the SDSS-III Collaboration.
\end{acknowledgements}

\begin{deluxetable}{lcrrrllcccl}
\footnotesize
\tablewidth{0pt}
\tablecaption{Comet-Like Main-Belt Asteroids\label{orbelems}}
\tablecolumns{11}
\tablehead{
\colhead{Name} & \colhead{Type\tablenotemark{a}} &
 \colhead{$a$\tablenotemark{b}} &
 \colhead{$e$\tablenotemark{c}} &
 \colhead{$i$\tablenotemark{d}} &
 \colhead{$T_J$\tablenotemark{e}} &
 \colhead{$P$\tablenotemark{f}} &
 \colhead{$H_R$\tablenotemark{g,h}} &
 \colhead{$r_{n}$\tablenotemark{i}} &
 \colhead{$(A_d/A_N)_{\rm max}$\tablenotemark{j,h}} &
 \colhead{Discovery Date\tablenotemark{k}}
}
\startdata
133P/Elst-Pizarro    & MBC & 3.160 & 0.162 &  1.386 & 3.184 & 5.62 & $15.49\pm0.05$ [1] & 1.9   & 0.7 [8]  & 1996 Aug  7 [12] \\
238P/Read            & MBC & 3.165 & 0.253 &  1.266 & 3.153 & 5.63 & $19.05\pm0.05$ [2] & 0.4   & 21 [2]   & 2005 Oct 24 [13] \\
176P/LINEAR          & MBC & 3.194 & 0.194 &  0.238 & 3.166 & 5.71 & $15.10\pm0.05$ [1] & 2.0   & 0.3 [9]  & 2005 Nov 26 [14] \\
P/2008 R1 (Garradd)  & MBC & 2.726 & 0.342 & 15.903 & 3.217 & 4.50 & $20.3\pm0.1$ [3]   & 0.2   & 28 [3]   & 2008 Sep  2 [15] \\
P/2010 A2 (LINEAR)   & DA  & 2.291 & 0.124 &  5.255 & 3.583 & 3.47 & $21.9\pm0.1$ [4]   & 0.06  & ---      & 2010 Jan  6 [16] \\
P/2010 R2 (La Sagra) & MBC & 3.099 & 0.154 & 21.395 & 3.099 & 5.46 & $17.9\pm0.2$ [5]   & 0.7   & 60 [5]   & 2010 Sep 14 [17] \\
(596) Scheila        & DA  & 2.928 & 0.165 & 14.661 & 3.209 & 5.01 &  8.54 [6]          & 56.67 & 2.2 [10] & 2010 Dec 10 [18] \\
P/2006 VW$_{139}$    & MBC & 3.052 & 0.201 &  2.438 & 3.203 & 5.33 & 15.9 [7]           & 1.7   & 1.1 [11] & 2011 Nov 05 [19]
\enddata
\tablenotetext{a}{Object classification as main-belt comet (MBC) or disrupted asteroid (DA).}
\tablenotetext{b}{Osculating semimajor axis, in AU.}
\tablenotetext{c}{Osculating eccentricity.}
\tablenotetext{d}{Osculating inclination, in degrees.}
\tablenotetext{e}{Tisserand parameter.}
\tablenotetext{f}{Orbital period, in years.}
\tablenotetext{g}{Absolute $R$-band magnitude of nucleus, assuming solar colors, in the IAU $H,G$ system.}
\tablenotetext{h}{References: [1] \citet{hsi09a}; [2] \citet{hsi11b}; [3] \citet{mac12}; [4] \citet{jew10}; [5] this work; [6] \citet{ted04};
                   [7] JPL Small Body Database ({\tt http://ssd.jpl.nasa.gov/sbdb.cgi}); [8] \citet{hsi10b}; [9] \citet{hsi11a}; [10] \citet{jew11}; [11] \citet{hsi12b};
                   [12] \citet{els96}; [13] \citet{rea05}; [14] \citet{hsi06a}; [15] \citet{gar08}; [16] \citet{bir10}; [17] \citet{nom10}; [18] \citet{lar10}; [19] \citet{hsi11c}.}
\tablenotetext{i}{Estimated effective radius of nucleus, from same works referenced for $H_R$, in km.}
\tablenotetext{j}{Peak observed dust-to-nucleus scattering surface area ratio.}
\tablenotetext{k}{Discovery date of comet-like activity.}
\end{deluxetable}

\begin{deluxetable}{lcccrrcccrrrrr}
\footnotesize
\tablewidth{0pt}
\tablecaption{Observation Log\label{obslog}}
\tablecolumns{15}
\tablehead{
\colhead{UT Date} & \colhead{Telescope\tablenotemark{a}} & \colhead{Moon\tablenotemark{b}}
  & \colhead{Seeing\tablenotemark{c}} & \colhead{N\tablenotemark{d}}
  & \colhead{$t$\tablenotemark{e}}  & \colhead{Filter}
  & \colhead{$R$\tablenotemark{f}} & \colhead{$\Delta$\tablenotemark{g}}
  & \colhead{$\alpha$\tablenotemark{h}}
  & \colhead{$\nu$\tablenotemark{i}}
  & \colhead{P.A.$_{-\odot}$\tablenotemark{j}}
  & \colhead{P.A.$_{-v}$\tablenotemark{k}}
  & \colhead{$\alpha_{\rm pl}$\tablenotemark{l}}
}
\startdata
2010 Jun 26     & {\it Perihelion} & --- & --- & --- & --- & --- & 2.623 & 2.228 & 22.3 &   0.0 & 241.2 & 242.8 &  -0.5 \\
2010 Aug 16     & PS1    & N$+$6  & 1.0 &  2 &   80 & $r'$   & 2.631 & 1.793 & 15.0 & 12.9 & 203.3 & 243.7 &  -9.4 \\
2010 Sep 08     & PS1    & N      & 0.9 &  1 &   40 & $r'$   & 2.641 & 1.739 & 12.0 & 18.5 & 162.1 & 241.1 & -11.7 \\
2010 Sep 19 (1) & Dk1.54 & N$+$11 & 1.4 & 20 & 2400 & R      & 2.649 & 1.756 & 12.3 & 21.3 & 138.8 & 239.7 & -12.0 \\
2010 Sep 19 (2) & UH2.2  & N$+$11 & 1.6 &  5 & 1500 & R      & 2.649 & 1.756 & 12.3 & 21.3 & 138.0 & 239.7 & -12.0 \\
2010 Sep 29     & FTN    & N$-$9  & 1.5 &  3 &  450 & R      & 2.652 & 1.790 & 13.4 & 23.7 & 119.9 & 238.7 & -11.7 \\
2010 Oct 05     & Keck   & N$-$2  & 0.9 &  2 &  240 & BR     & 2.659 & 1.824 & 14.4 & 25.2 & 110.7 & 238.2 & -11.3 \\
2010 Oct 05     & Keck   & N$-$2  & 0.9 &  4 & 3600 & Spec.  & 2.659 & 1.824 & 14.4 & 25.2 & 110.7 & 238.2 & -11.3 \\
2010 Oct 19     & UH2.2  & N$+$11 & 1.1 &  3 &  900 & R      & 2.666 & 1.922 & 16.8 & 28.5 &  94.5 & 237.7 &  -9.8 \\
2010 Nov 26     & UH2.2  & N$-$9  & 1.8 &  6 & 1800 & R      & 2.698 & 2.343 & 21.1 & 37.8 &  71.3 & 239.6 &  -4.0 \\
2010 Dec 12     & UH2.2  & N$+$6  & 0.8 & 21 & 6300 & R      & 2.713 & 2.535 & 21.3 & 41.3 &  65.7 & 241.4 &  -1.5 \\ 
2011 Dec 31     & INT    & N$-$3  & 1.1 &  4 & 1200 & R      & 2.733 & 2.793 & 20.5 & 45.9 &  60.5 & 243.9 &   1.2 \\
2011 Feb 03     & Keck   & N      & 0.8 &  2 &  360 & BR     & 2.772 & 3.210 & 17.0 & 53.5 &  53.4 & 248.7 &   4.4 \\
2011 Aug 26     & Keck   & N$-$3  & 1.0 &  4 & 1200 & BR     & 3.066 & 3.289 & 17.9 & 95.1 & 270.9 & 291.6 &  -5.7 \\
2011 Aug 31     & Gemini & N$+$2  & 0.8 &  6 & 1800 & $r'$   & 3.073 & 3.231 & 18.2 & 96.0 & 270.8 & 292.5 &  -6.0 \\
2013 Mar 13     & {\it Aphelion}   & --- & --- & --- & --- & --- & 3.570 & 2.758 & 10.5 & 180.0 &  76.6 & 302.8 &  7.5 \\
2015 Nov 30     & {\it Perihelion} & --- & --- & --- & --- & --- & 2.620 & 2.946 & 19.3 &   0.0 &  67.0 & 238.0 & -2.8  
\enddata
\tablenotetext{a} {Telescope used (PS1:\ 1.8-m Pan-STARRS 1 telescope; Dk1.54:\ Danish 1.54-m telescope; FTN:\ 2.0-m Faulkes Telescope North;
                    Gemini: 8-m Gemini-North telescope; Keck:\ Keck I 10-m telescope; UH2.2:\ University of Hawaii 2.2-m telescope;
                    INT:\ 2.5-m Isaac Newton Telescope)}
\tablenotetext{b} {Phase of moon expressed in offset from new moon (``N'')
                    in days.}
\tablenotetext{c} {Approximate average seeing (FWHM) in arcsec.}
\tablenotetext{d} {Number of images.}
\tablenotetext{e} {Total effective exposure time in seconds.}
\tablenotetext{f} {Heliocentric distance in AU.}
\tablenotetext{g} {Geocentric distance in AU.}
\tablenotetext{h} {Solar phase angle (Sun - P/La Sagra - Earth) in degrees.}
\tablenotetext{i} {True anomaly in degrees.}
\tablenotetext{j} {Position angle of the antisolar vector, as projected in the plane of the sky, in degrees east of north.}
\tablenotetext{k} {Position angle of the negative velocity vector, as projected in the plane of the sky, in degrees east of north.}
\tablenotetext{l} {Orbit-plane angle (between observer and object orbit plane as seen from object) in degrees.}
\end{deluxetable}

\begin{deluxetable}{lcccccccc}
\footnotesize
\tablewidth{0pt}
\tablecaption{Photometry Results\label{photresults}}
\tablecolumns{11}
\tablehead{
\colhead{UT Date} & \colhead{$m_{\rm avg}$\tablenotemark{a}}
  & \colhead{$m_{\rm avg}(1,1,0)$\tablenotemark{b}}
  & \colhead{$m_{\rm tot}$\tablenotemark{c}}
  & \colhead{$m_{\rm tot}(1,1,0)$\tablenotemark{d}}
  & \colhead{$A_{d}/A_{N}$\tablenotemark{e}}
  & \colhead{$A_{d}$\tablenotemark{f}}
  & \colhead{$M_{d}$\tablenotemark{g}}
  & \colhead{$Af\rho$\tablenotemark{h}}
}
\startdata
2010 Aug 16     & $18.49\pm0.05$ & $14.25\pm0.15$ & $18.3\pm0.1$   & $14.0\pm0.2$ & $30\pm10$  & $40\pm15$  & $7\pm2$  & $42\pm6$ \\
2010 Sep 08     & $18.26\pm0.05$ & $14.20\pm0.15$ & $18.0\pm0.1$   & $13.9\pm0.2$ & $30\pm10$  & $45\pm15$  & $7\pm2$  & $45\pm6$ \\
2010 Sep 19 (1) & $18.28\pm0.02$ & $14.15\pm0.15$ & $17.9\pm0.1$   & $13.8\pm0.2$ & $30\pm10$  & $45\pm15$  & $8\pm2$  & $46\pm6$ \\
2010 Sep 19 (2) & $18.23\pm0.02$ & $14.10\pm0.15$ & $17.8\pm0.1$   & $13.7\pm0.2$ & $30\pm10$  & $45\pm15$  & $8\pm2$  & $48\pm6$ \\
2010 Sep 29     & $18.39\pm0.02$ & $14.19\pm0.15$ & $18.0\pm0.1$   & $13.8\pm0.2$ & $30\pm10$  & $45\pm15$  & $8\pm2$  & $44\pm6$ \\
2010 Oct 05     & $18.34\pm0.03$ & $14.05\pm0.20$ & $18.0\pm0.1$   & $13.7\pm0.2$ & $35\pm10$  & $50\pm20$  & $8\pm3$  & $49\pm7$ \\
2010 Oct 19     & $18.47\pm0.03$ & $13.97\pm0.20$ & $18.1\pm0.1$   & $13.6\pm0.2$ & $35\pm10$  & $55\pm20$  & $9\pm3$  & $50\pm8$ \\
2010 Nov 26     & $18.89\pm0.03$ & $13.80\pm0.20$ & $18.4\pm0.1$   & $13.3\pm0.2$ & $45\pm15$  & $65\pm20$  & $11\pm4$ & $49\pm9$ \\
2010 Dec 12     & $18.78\pm0.02$ & $13.50\pm0.20$ & $18.4\pm0.1$   & $13.1\pm0.2$ & $60\pm20$  & $85\pm30$  & $14\pm5$ & $60\pm11$ \\ 
2010 Dec 31     & $18.93\pm0.04$ & $13.45\pm0.20$ & $18.2\pm0.2$   & $12.7\pm0.2$ & $60\pm20$  & $85\pm30$  & $15\pm5$ & $57\pm10$\tablenotemark{i} \\
2011 Feb 03     & $19.53\pm0.08$ & $13.85\pm0.20$ & $19.1\pm0.1$   & $13.4\pm0.2$ & $40\pm10$  & $60\pm20$  & $10\pm3$ & $34\pm6$ \\
2011 Aug 26     & $24.1\pm0.1$   & $18.1\pm0.3$   & ---            & ---          & ---        & ---        & ---      & $0.7\pm0.1$ \\
2011 Aug 31     & $23.4\pm0.1$   & $17.4\pm0.3$   & ---            & ---          & ---        & ---        & ---      & $1.2\pm0.2$ 
\enddata
\tablenotetext{a}{Mean apparent magnitude inside a $5\farcs0$ photometry aperture.}
\tablenotetext{b}{Mean magnitude, normalized to $R=\Delta=1$~AU and $\alpha=0\degr$, assuming $-0.03<G_R<0.26$.}
\tablenotetext{c}{Total apparent magnitude inside rectangular aperture enclosing entire comet (nucleus and dust tail).}
\tablenotetext{d}{Total magnitude, normalized to $R=\Delta=1$~AU and $\alpha=0\degr$, assuming $-0.03<G_R<0.26$.}
\tablenotetext{e}{Inferred ratio of scattering surface area of dust to nucleus scattering surface area.}
\tablenotetext{f}{Inferred scattering surface area of dust, in $10^6$~m$^2$, using $r_N=0.6$~km.}
\tablenotetext{g}{Estimated dust mass, in $10^5$~kg, assuming 10~$\mu$m-radius grains and $\rho=1300$~kg~m$^{-3}$.}
\tablenotetext{h}{Dust contribution (computed using a $5\farcs0$ photometry aperture), as parameterized by \citet{ahe84}, in cm.}
\tablenotetext{i}{Computed from data in which comet was trailed by $4\farcs1$, and as such, represents a lower limit.}
\end{deluxetable}


\begin{deluxetable}{llllll}
\tablewidth{0pt}
\tablecaption{P/La Sagra Cluster Members\label{lasagcluster}}
\tablecolumns{6}
\tablehead{
\colhead{Object}
 & \colhead{$a_p$\tablenotemark{a}}
 & \colhead{$e_p$\tablenotemark{b}}
 & \colhead{$\sin(i_p)$\tablenotemark{c}}
 & \colhead{$H_V$\tablenotemark{d}}
 & \colhead{$T_{lyap}$\tablenotemark{e}}
}
\startdata
18901 (2000 MR$_{5}$)    & 3.0942 & 0.1223 & 0.3791 & 12.1 & 683 \\
106020 (2000 SS$_{294}$) & 3.0928 & 0.1146 & 0.3812 & 14.4 & 480 \\
106064 (2000 SA$_{323}$) & 3.0931 & 0.1215 & 0.3824 & 14.5 & 781 \\
131501 (2001 SX$_{272}$) & 3.0889 & 0.1177 & 0.3831 & 14.2 & 999 \\
       (2002 RQ$_{261}$) & 3.0887 & 0.1150 & 0.3824 & 15.4 & 900 \\
P/2010 R2 (La Sagra)     & 3.1001 & 0.1148 & 0.3817 & 18.3 & 270
\enddata
\tablenotetext{a}{Proper semimajor axis, in AU}
\tablenotetext{b}{Proper eccentricity}
\tablenotetext{c}{Sine of proper inclination}
\tablenotetext{d}{Absolute magnitude, in $V$-band}
\tablenotetext{e}{Lyapunov time, in kyr}
\end{deluxetable}



\begin{figure}
\plotone{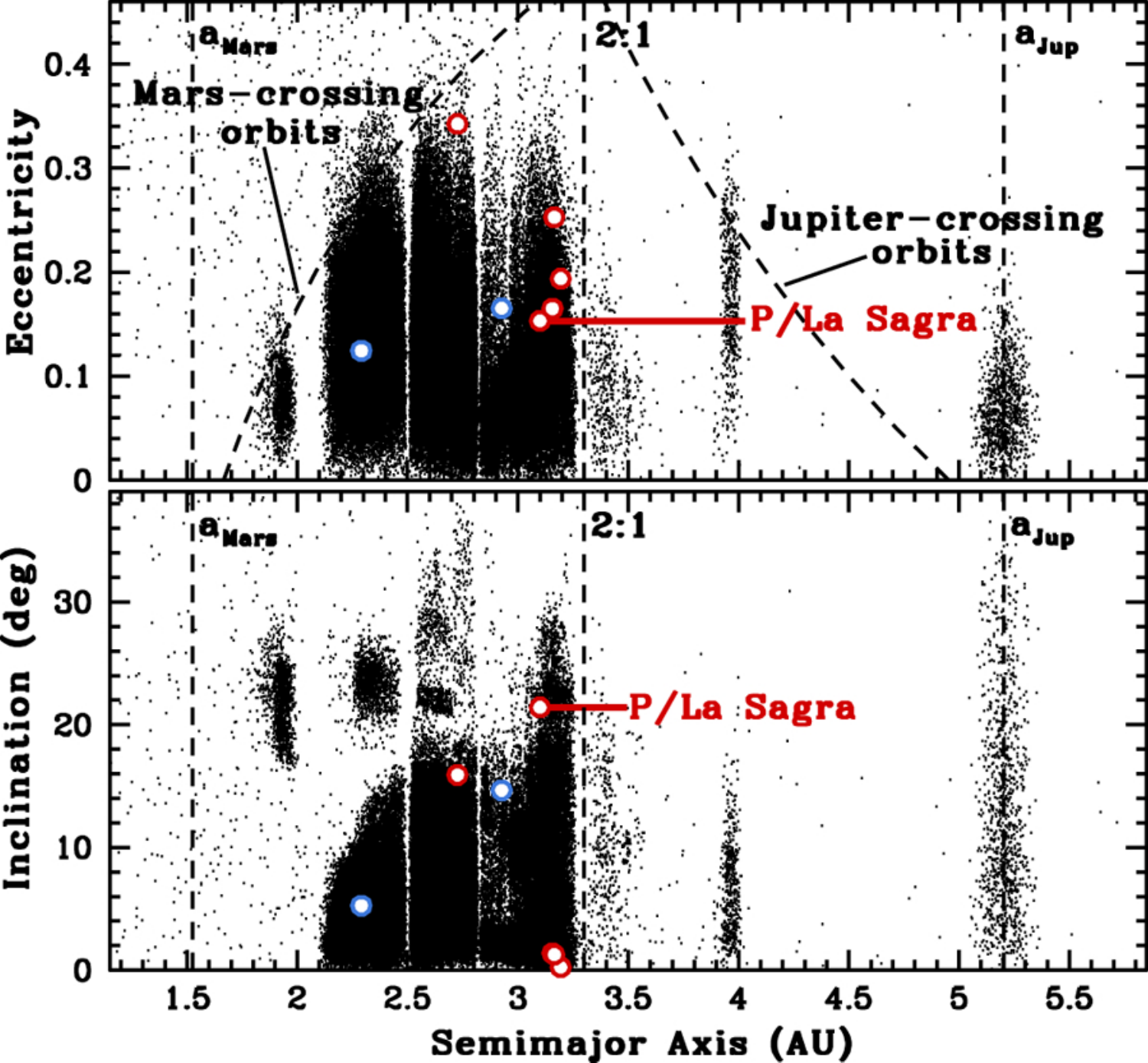}
\caption{\small Plots of eccentricity (upper panel) and inclination (lower panel) versus semimajor axis
showing the distributions in orbital element space of main-belt asteroids (black dots),
MBCs (red circles), and likely disrupted asteroids (blue circles). Also marked with dotted
lines are the semimajor axes of Mars ($a_{\rm Mars}$) and Jupiter ($a_{\rm Jup}$),
the semimajor axis of the 2:1 mean-motion resonance with Jupiter, and the loci
of Mars-crossing orbits and Jupiter-crossing orbits.
}
\label{aeiplots}
\end{figure}

\begin{figure}
\plotone{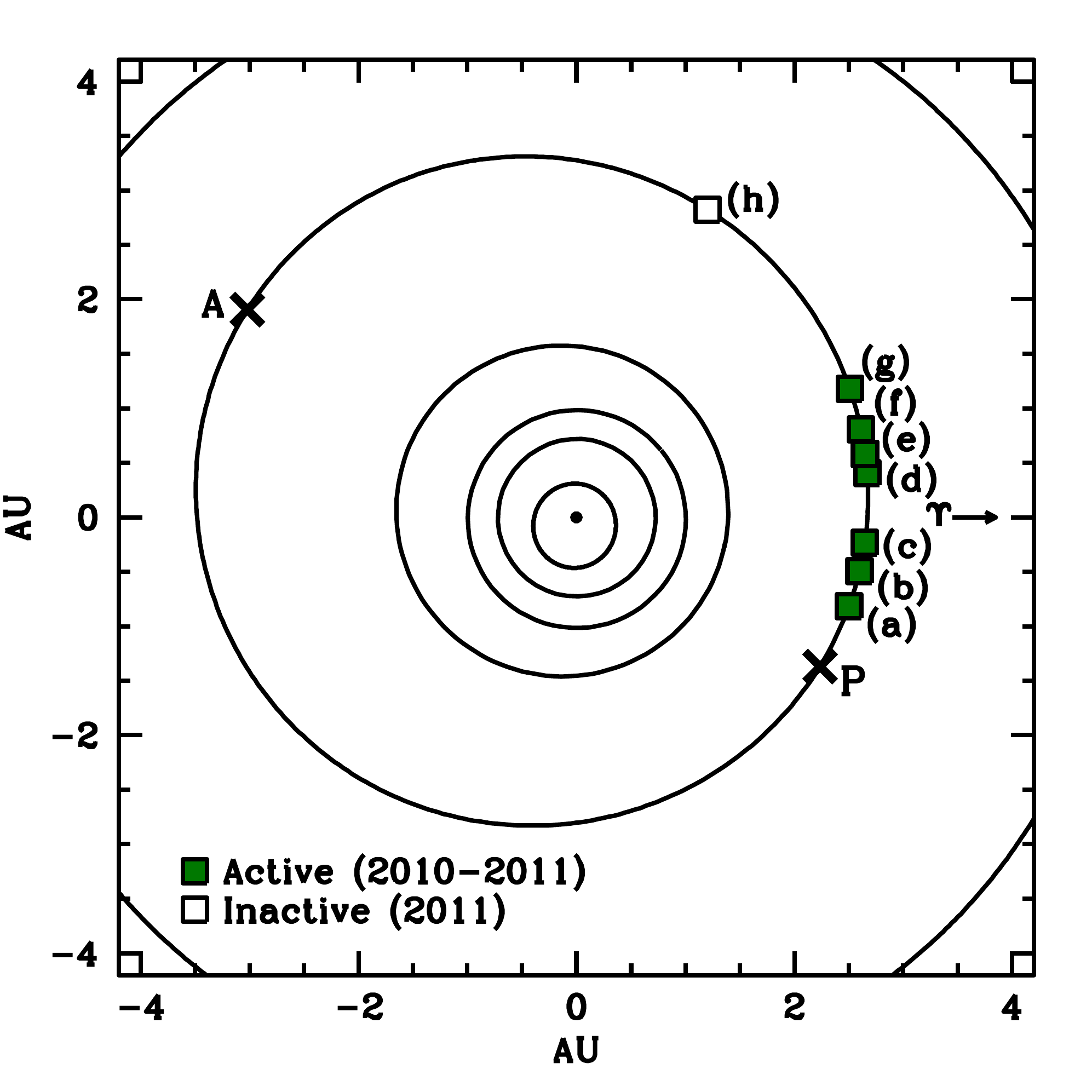}
\caption{Orbital position plot of P/La Sagra observations detailed in Table~\ref{obslog}.
The Sun is shown at the center as a solid dot, with the orbits of Mercury, Venus, Earth,
Mars, P/La Sagra, and Jupiter (from the center of the plot outwards) shown as black lines.
Solid squares mark positions where P/La Sagra was observed to be active in 2010 and early 2011,
while an open square marks the position where P/La Sagra was observed to be inactive in August 2011.
Perihelion (P) and aphelion (A) positions are marked with crosses.
Observations plotted were obtained on (a) 2010 August 16, (b) 2010 September 8-19, (c) 2010 October 5,
(d) 2010 November 25-28, (e) 2010 December 12, (f) 2010 December 31, (g) 2011 February 3,
and (h) 2011 August 26-31.
}
\label{actvplot}
\end{figure}



\end{document}